\documentclass[twoside]{article}

\usepackage{PRIMEarxiv}

\usepackage[utf8]{inputenc} % allow utf-8 input
\usepackage[T1]{fontenc}    % use 8-bit T1 fonts
\usepackage{hyperref}       % hyperlinks
\usepackage{url}            % simple URL typesetting
\usepackage{booktabs}       % professional-quality tables
\usepackage{amsfonts}       % blackboard math symbols
\usepackage{nicefrac}       % compact symbols for 1/2, etc.
\usepackage{microtype}      % microtypography
\usepackage{fancyhdr}       % header
\usepackage{graphicx}       % graphics
\usepackage{rotating}
\usepackage{longtable, array}
\usepackage{tabularx}
\usepackage{pdflscape}
\usepackage{amssymb,amsthm,amsmath}
\usepackage{adjustbox}
\usepackage{placeins}

\usepackage{natbib}
\usepackage{subfigure}

\usepackage{stackengine}
\newcommand\xrowht[2][0]{\addstackgap[.5\dimexpr#2\relax]{\vphantom{#1}}}

\newbox\subfigbox
\makeatletter
\newenvironment{subfloat}
{\def\caption##1{\gdef\subcapsave{\relax##1}}%
\let\subcapsave\@empty
\setbox\subfigbox\hbox
\bgroup}
{\egroup
\subfigure[\subcapsave]{\box\subfigbox}}
\makeatother

\title{The Controlled Four-Parameter Method for Cross-Assignment of Directional Wave Systems
}

\author{
  Andre Luiz Cordeiro dos Santos \\
  Mathematics Institute, Federal Center for Technological Education of Rio de Janeiro (CEFET-RJ), Brazil \\  
  \texttt{andre.santos@cefet-rj.br} \\
   \And
  Felipe Marques dos Santos\\
  National Oceanography Centre, Southampton, UK\\  
  \texttt{felipe.santos@noc.ac.uk} \\
   \AND
   Nelson Violante-Carvalho \\
   Ocean Engineering Program, Rio de Janeiro Federal University (COPPE-UFRJ), Brazil \\  
   \texttt{n\_violante@oceanica.ufrj.br} \\
    \And
   Luiz Mariano Carvalho\\
   Applied Mathematics Department, Institute of Mathematics and Statistics\\ Rio de Janeiro State University (UERJ), Brazil\\
   \texttt{luizmc@ime.uerj.br} \\
   \AND  
   Helder Manoel Venceslau \\
   Mathematics Institute, Federal Center for Technological Education of Rio de Janeiro (CEFET-RJ), Brazil \\  
   \texttt{helder.venceslau@cefet-rj.br} \\  
}

\begin{document}
\maketitle

\begin{abstract}
Cross-assignment of directional wave spectra is a critical task in wave data assimilation. Traditionally, most methods rely on two-parameter spectral distances or energy ranking approaches, which often fail to account for the complexities of the wave field, leading to inaccuracies. To address these limitations, we propose the Controlled Four-Parameter Method (C4PM), which independently considers four integrated wave parameters. This method enhances the accuracy and robustness of cross-assignment by offering flexibility in assigning weights and controls to each wave parameter. We compare C4PM with a two-parameter spectral distance method using data from two buoys moored 13 km apart in deep water. Although both methods produce negligible bias and high correlation, C4PM demonstrates superior performance by preventing the occurrence of outliers and achieving a lower root mean square error across all parameters. The negligible computational cost and customization make C4PM a valuable tool for wave data assimilation, improving the reliability of forecasts and model validations.
\end{abstract}

% keywords can be removed

\section{Introduction}

The wave directional spectrum is the fundamental representation of a sea state, providing a detailed description of the energy distribution as a function of both frequency and direction~---~hence crucial for understanding the complexities of wave dynamics and interactions. 
Accurate analysis of the spectrum is vital for various applications, including climate studies, coastal management and maritime safety \citep{luigi, 10.3389/fmars.2019.00124}. 
One of the key challenges is the cross-assignment of spectral partitions, which involves identifying and matching collocated wave systems from different datasets or models~---~see a discussion about spectral partitioning in \cite{gerling, nvc2, jesus, jesus2}. 
Cross-assignment is essential for data assimilation, where observational data are integrated into numerical models \citep{Aouf, 9104702}, and for assessing measurements, enabling comparison of results from various sources.

%Traditionally, the cross-assignment of spectral partitions relies heavily on evaluating the spectral distance in the frequency-direction ($f-\theta$) or wavenumber ($k_x - k_y$) space. In \cite{hasselmann1996}, an expression was proposed that represents the normalized square distance in wavenumber space used for cross-assigning wave systems in the form%

Currently, most (if not all) methods for cross-assignment are based either on two-parameter spectral distance or on ranking the energy content of each system~---~the most energetic partitions in one spectrum are paired to the correspondingly ranked partitions in the other spectrum.
Energy ranking methods are more prone to inaccuracies, mainly when the number of partitions of each spectra differ.
Cross-assigning only the most energetic partition might overcome this limitation but significantly reduce the possible number of matches. 
Conversely, methods based on a two-parameter spectral distance rely exclusively on frequency and direction. % with different flavors expressed by Equation~\eqref{hasselmann} 
\citep[see, among many others,][]{hasselmann1996, hanson, LI201272, wang, SMIT2021101738, 10.1029/2020GL091187, felipe, 9537908,  RICONDO2023102210, WU2024102397}.
However, this approach has drawbacks. 
The main limitation is that it can result in errors, mainly because %the method disregards any energy discrepancy or when 
partitions close in frequency but significantly apart in direction (or vice versa) are mismatched~---~leading to potential discrepancies caused by outliers. 
These inaccuracies can propagate through data assimilation processes, resulting in suboptimal model performance and potentially misleading data interpretations.

To overcome these limitations, we propose a novel methodology for cross-assigning partitions, termed the Controlled Four-Parameter Method (C4PM). This approach incorporates four bulk wave spectral parameters: significant wave height, peak wave period, peak wave direction, and peak wave spreading. By independently treating these parameters, the method enhances the robustness and precision of the cross-assignment process, positively affecting the assimilation of observational data into numerical models and thereby improving the forecast results. C4PM has proven to be an valuable and efficient tool to cross-assignment, allowing to define control levels for each of the primary wave parameters and additionally prioritize parameters by assigning weights through a customizable weighting vector.

The structure of the current analysis is organized as follows: Section~\ref{back} presents the theoretical framework of C4PM, which includes the definition of the semimetric on which the method is based, as well as a detailed presentation of the cross-assignment scheme used.
Section~\ref{MET} outlines the buoy data and the processing methodologies employed, while Section~\ref{res} discusses the results. 
Finally, the conclusions are summarized in Section~\ref{SC}.

%\section{Analytical Background}

\section{The Cross-Assignment Problem}\label{back}

%This section addresses the cross-assignment problem, starting with its basic concept, followed by a quick description of the 2PM technique and its limitations, and, finally, a thorough description of the C4PM method.
This section addresses the cross-assignment problem, beginning with a basic concept, followed by a brief description of a classic technique and its limitations, and concluding with a detailed explanation of the C4PM method.

\subsection{Matching: a fundamental concept}

Assuming that \( A(\mathbf{k}) \) and \( B(\mathbf{k}) \) are two measurements of the wavenumber spectrum associated with a certain sea state. After a partitioning process of \( A(\mathbf{k}) \) and \( B(\mathbf{k}) \), consider the additive decomposition:

\begin{equation}
    \left\{\begin{array}{l}
  A (\mathbf{k}) = A_1 (\mathbf{k}) + A_2 (\mathbf{k}) + \cdots + A_p
  (\mathbf{k})\\
  B (\mathbf{k}) = B_1 (\mathbf{k}) + B_2 (\mathbf{k}) + \cdots + B_m
  (\mathbf{k})
\end{array}\right.,
\end{equation} where \( A_{1}(\mathbf{k}), A_{2}(\mathbf{k}), \ldots, A_{p}(\mathbf{k}) \) and \( B_{1}(\mathbf{k}), B_{2}(\mathbf{k}), \ldots, B_{m}(\mathbf{k}) \) are, respectively, the partitions\footnote{We use the term "partition" to refer to an independent wave system that is a constituent of the wave spectrum} of \(A(\mathbf{k})\) and \( B(\mathbf{k}) \). It is supposed from now on that \( p \leqslant m \) and that the explicit dependence on the wavenumber vector \(\mathbf{k}\) will be omitted.

We call \(n\)-\textit{matching} between \(A\) and \(B\) a set

\begin{equation}\label{matching}
\mathcal{M} = \left\{\left\{A_{i_1}, B_{j_1}\right\} ;\left\{A_{i_2}, B_{j_2}\right\} ; \ldots ;\left\{A_{i_n}, B_{j_n}\right\}\right\},
%\begin{equation}\label{matching}
%\mathcal{M} = \left\{\left\{A_{k_1}, B_{j_{k_1}}\right\} ;\left\{A_{k_2}, B_{j_{k_2}}\right\} ; \ldots ;\left\{A_{k_n}, B_{j_{k_n}}\right\}\right\},
\end{equation} formed by \(n \leqslant p\) of pairs\footnote{unordered} of partitions of \(A\) and \(B\) with the property that no partition of \(A\) is associated with more than one partition of \(B\) and vice versa. 
%Ideally, a cross-assignment is a matching between \(A\) and \(B\) such that matched partitions should have oceanic characteristics with the highest possible concordance. Thereby, it is essential to define a technique capable of identifying compatible wave systems with the greatest possible accuracy.
Ideally, a cross-assignment is a matching between \(A\) and \(B\) such that coupled partitions exhibit the highest possible concordance in their oceanic characteristics. 
Therefore, it is crucial to employ a technique capable of accurately identifying compatible wave systems.
%formed by \(n \leqslant p\) of unordered pairs of partitions of \(A\) and \(B\) with the property that if \(1\leqslant i \neq i'\leqslant n\) then \(1\leqslant k_i\neq k_{i'}\leqslant p\) and \( 1 \leqslant j_{k_{i}} \neq j_{k_{i'}} \leqslant m\). This condition simply ensures that a partition of \(A\) is not associated with more than one partition of \(B\), and vice-versa.

%A \textit{cross-assignment} between \(A\) and \(B\) is a \(p\)-matching, i.e., a set
%\begin{equation}\label{pareamento}
%\mathcal{P} = \left\{\left(A_{1}, B_{j_{1}}\right) ;\left(A_{2}, B_{j_{2}}\right) ; \ldots ;\left(A_{p}, B_{j_{p}}\right)\right\} ,   
%\end{equation}
%formed by \(p\) pairs of partitions of \(A\) and \(B\) with the property that \textit{if \( 1 \leqslant i \neq i' \leqslant p \) then \( 1 \leqslant j_{i} \neq j_{i'} \leqslant m \)}.

\subsection{A two-parameter method (2PM)}\label{2PM}

\begin{figure}[ht!]    
    \centering
    \begin{subfloat}{
        \centering
        \includegraphics[width=13cm]{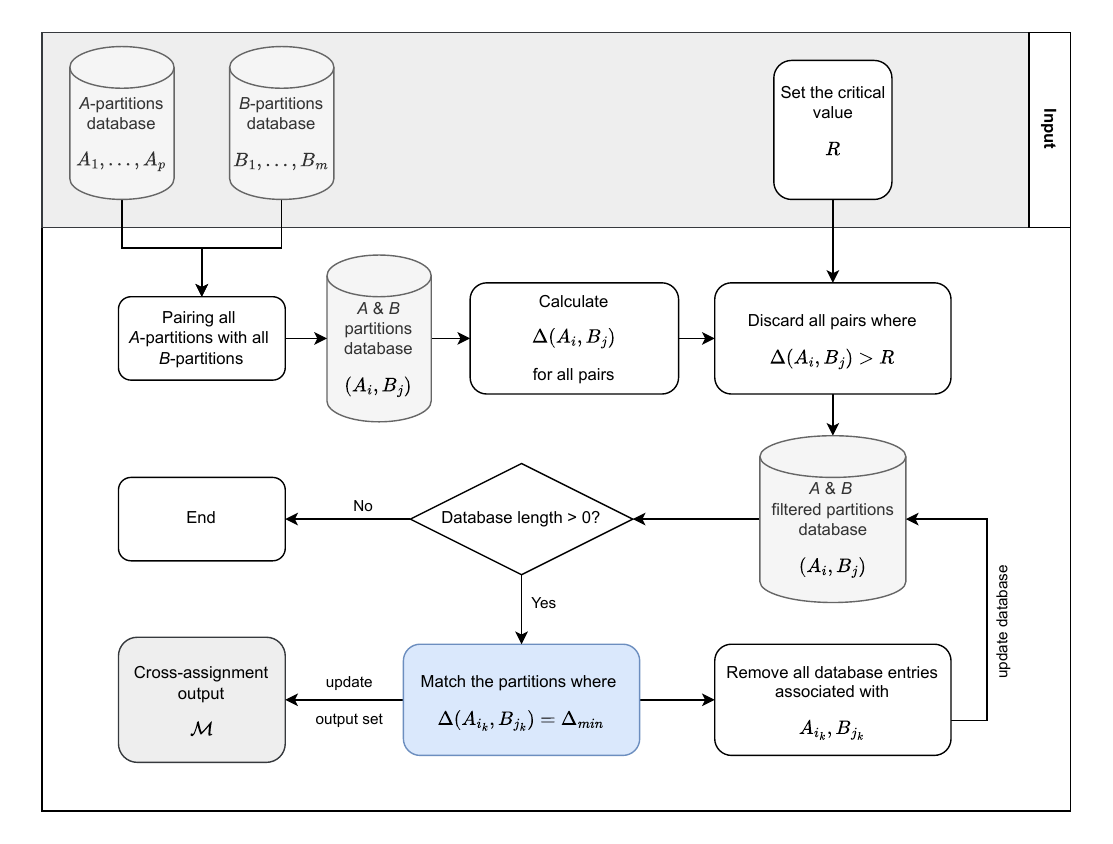}}
    \end{subfloat}
    \caption{2PM algorithm diagram.}\label{fig:2PMdiag}
\end{figure}

Typically, the cross-assignment of spectral partitions heavily relies on evaluating a two-parameter spectral distance. In this context, the fundamental concept behind the technique proposed in \cite{hasselmann1996} is the association of a partition with its so-called (two-parameter) characteristic wavenumber vector. In that study, the expression
\begin{equation}\label{hasselmann}
    \Delta(A_i,B_j) = \frac{\|\mathbf{k}(A_{i}) - \mathbf{k}(B_j)\|}{\sqrt{\|\mathbf{k}(A_{i})\|^2 + \|\mathbf{k}(B_{j})\|^2}},
\end{equation} where \(\mathbf{k}(A_{i})\) and \(\mathbf{k}(B_j)\) are the characteristic wavenumber vectors of \( A_i \) and \( B_j \), respectively, was proposed for the distance\footnote{the double-bar symbol designates the Euclidean norm} between partitions of \( A \) and \( B \). 
Thus, the \(n\)-matching given by \eqref{matching} is the cross-assignment between $A$ and $B$, in the Hasselmann sense, if the following conditions are met:
\begin{enumerate}
    \item[(i)] \(n\leqslant p\) is the largest possible \vskip0.5cm 
    \item[(ii)]  \(\Delta(A_{i_1},B_{j_1}) \leqslant \Delta(A_{i_2},B_{j_2}) \leqslant \ldots \Delta(A_{i_n},B_{j_n}) \leqslant R\) \vskip0.5cm
    \item[(iii)] \(\Delta(A_{i_k},B_{j_k}) = \min \{\Delta(A_{i_k},B_{j}): j=1,2,\ldots,m\}\) for each \(k=1,2,\ldots,n\).
\end{enumerate} Here, \(R\) is a specified critical value, which acts as a cutoff line defining which pairs will be in the cross-assignment. 
%The method computes all possible distances \(\Delta(A_i,B_j)\), sort the remaining pairs based on those values, removes all pairs with distances greater than a specified critical value (\(R\) nominally 0.75), and then, finally, sets the partition matchups, starting from the smallest distances. The largest matching thus formed is the cross-assignment between the partitions of \(A\) and \(B\) wavenumber spectra according to \cite{hasselmann1996}.
%We will refer to this technique from now on as 2PM; its algorithm is presented in Figure \ref{fig:2PMdiag}. The method computes all possible distances between partitions of \(A\) and \(B\), removes all pairs with distances greater than the value of \(R\); finally, sets the partition matchups, starting from the smallest distances. %That work also suggests that satisfactory results were achieved using $R = 0.75$ as the method's critical value.
Hereafter, we will refer to this technique as 2PM, and its algorithm is illustrated in Figure~\ref{fig:2PMdiag}. 
The method calculates all possible distances between the partitions of \(A\) and \(B\), eliminates pairs with distances exceeding the threshold value \(R\), and finally establishes partition matchups by prioritizing the smallest distances.

\begin{figure}[ht!]
    \centering
    \begin{subfloat}{
        \centering
        \includegraphics[width=6cm]{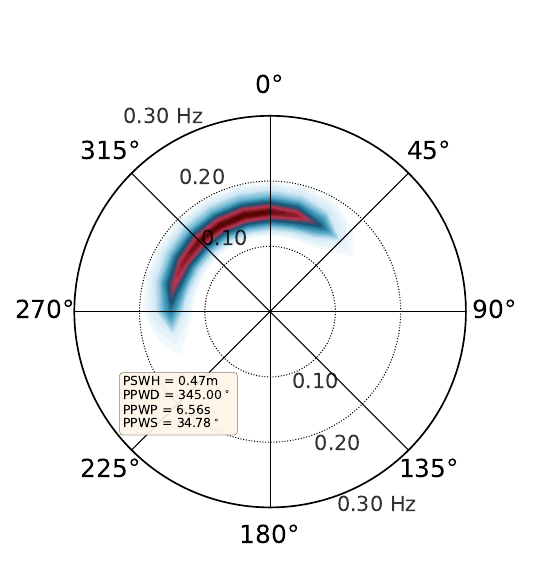}
        \label{subfig:bad_ca1_a}}
        \caption{}
    \end{subfloat}
    %\hfill
    \hspace*{0.5cm}
    \begin{subfloat}{
        \centering
        \includegraphics[width=6cm]{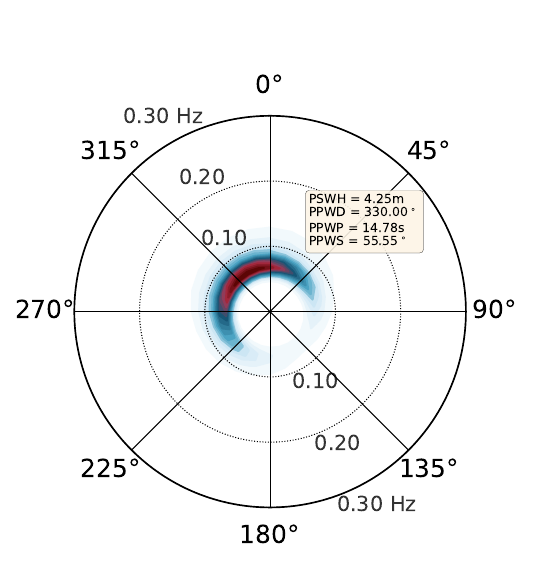}
        \label{subfig:bad_ca1_b}}
        \caption{}
    \end{subfloat} 
\caption{Directional wave spectra from NDBC buoys: (a) 51001 and (b) 51101 on 12 Jan 2023 at 18:40:00. The boxes indicate  the values of the partitioned integrated wave parameters~---~partitioned significant wave height ($PSWH$), partitioned peak wave period ($PPWP$), partitioned peak wave direction ($PPWD$) and partitioned peak directional spreading ($PPWS$). The \(\Delta\)-distance between partitions is~\(0.59\). }
\label{fig:bad_ca_hh_1a}
\end{figure}

\begin{figure}[ht!]
    \centering
    \begin{subfloat}{
        \centering
        \includegraphics[width=6cm]{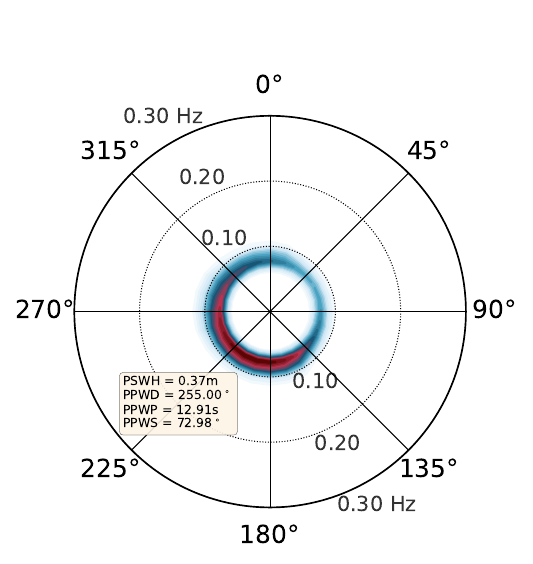}
        \label{subfig:bad_ca2_a}}
        \caption{}
    \end{subfloat}
    %\hfill
    \hspace*{0.5cm}
    \begin{subfloat}{
        \centering
        \includegraphics[width=6cm]{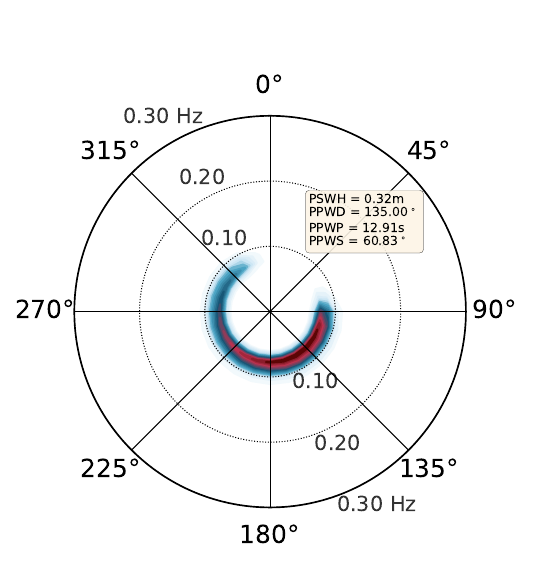}
        \label{subfig:bad_ca2_b}}
        \caption{}
    \end{subfloat}   
\caption{Directional wave spectra from NDBC buoys: (a) 51001 and (b) 51101 on 02 Aug 2023 at 01:10:00. The boxes indicate  the values of the partitioned integrated wave parameters. The \(\Delta\)-distance between partitions is \( 0.32\).}
\label{fig:bad_ca_hh_2a}
\end{figure}

% One of the main shortcomings of this approach is that, even if the spectral distance between two partitions is considered small~---~below a given $R$~---~it is not possible to guarantee that there is a good agreement between the wave systems characterized by them. 
% For example, in \cite{hasselmann1996} there is an indication that the critical value $0.75$ produces satisfactory results.
One of the main shortcomings of this approach is that even if the spectral distance between two partitions is considered small~---~below a given threshold ~---~it does not guarantee a strong agreement between the wave systems they represent. 
For instance, \cite{hasselmann1996} suggests that a critical value of $0.75$ yields satisfactory results.
%However, Figures~\ref{fig:bad_ca_hh_1a} and \ref{fig:bad_ca_hh_2a} present two examples of cross-assigned partitions whose distances are less than $R=0.75$, but their wave parameters have relevant discrepancies. 
However, Figures~\ref{fig:bad_ca_hh_1a} and \ref{fig:bad_ca_hh_2a} illustrate two examples of cross-assigned partitions with distances less than $0.75$, yet exhibiting significant discrepancies in their wave parameters.
In Figure \ref{fig:bad_ca_hh_1a}, the distance between the partitions is equal to~$0.59$, but the partitioned significant wave height is approximately nine times larger than its assigned counterpart, meaning that unrelated wave systems were associated in the cross-assignment process. 
A more subtle case of mismatch with distance equal to~0.32 is depicted in Figure \ref{fig:bad_ca_hh_2a}. 
Despite similar values of partitioned peak wave periods, partitioned significant wave heights and partitioned peak directional spreadings, the wave systems propagating towards the southern quadrant are separated by 120$^\circ$.

%The main shortcoming of this approach is that even if the spectral distance between two partitions is considered small, it is not possible to guarantee that there is a good agreement between the wave systems characterized by these components. In fact, Figures \ref{fig:bad_ca_hh_1a} and \ref{fig:bad_ca_hh_2a} bring two examples of poor cross-assigned partitions which ilustrate the main issues related to this technique.
%The first one shows matched partitions with very distinct peak periods and wave height, where $\Delta$=0.59, hence below the threshold value, indicating a potentially good match. The wave systems are unrelated, with the partition depicted in Figure~\ref{fig:bad_ca_hh_1a}b exhibiting a wave height approximately nine times greater than its assigned counterpart. The second represents a more subtle case of mismatch with $\Delta$=0.32. Despite similar values of period, wave height and spreading, the wave systems propagating towards the south quadrant are 120$^\circ$ apart.

\subsection{The Controlled Four-Parameter Method - C4PM}

%As shown in~\ref{2PM}, 2PM for cross-assignment has inexpedient limitations as it does not take into consideration all the main partitioned integrated wave parameters, leaving a significant margin for poor matches in the cross-assignment. In this sense, the \textit{Controlled Four Parameter Method} (C4PM) is being proposed as a more robust and efficient tool to handle the cross-assignment process with full control on the wave parameters variabilities so as on their hierarchy. 
%As shown in~\ref{2PM}, the 2PM method for cross-assignment has limitations, as it does not account for all primary partitioned integrated wave parameters, leaving a substantial margin for poor matches in the cross-assignment process. 
%In this context, the \textit{Controlled Four-Parameter Method} (C4PM) is proposed as a more robust and efficient tool, offering full control over wave parameter variabilities as well as their hierarchy.
As shown in~\ref{2PM}, 2PM methods for cross-assignment have significant limitations, thereby leaving margin for mismatches in the cross-assignment process. 
In this context, the \textit{Controlled Four-Parameter Method} (C4PM) is introduced as a more robust and efficient tool, providing comprehensive control over wave parameter variabilities and their hierarchical importance.

\subsubsection{The Weighted Semimetric}
%We begin by recalling the definition of a semimetric and proceed to concretely define our notion of spectral distance between two partitions. This spectral distance is based on a semimetric that allows the assignment of weights, reflecting the relative importance of each partitioned integrated wave parameter.

A semimetric defined on the partitions of \(A\) and \(B\) is a function \( d \) such that for any \( 1 \leqslant i \leqslant p \) and \( 1 \leqslant j \leqslant m \), the following conditions hold: \( d\left(A_{i}, B_{j}\right) \geqslant 0 \), \( d\left(A_{i}, B_{j}\right) = d\left(B_{j}, A_{i}\right) \), and \( d\left(A_{i}, B_{j}\right)=0 \) if, and only if, \( A_{i}=B_{j} \).
To define our spectral distance, each partition is associated with a four-dimensional vector whose components represent the partitioned integrated wave parameter values that characterize it: partitioned significant wave height ($PSWH$), partitioned peak wave period ($PPWP$), partioned peak wave direction ($PPWD$), and partioned peak wave spreading ($PPWS$), precisely in that order. 
This approach incorporates more information about each partition compared to the distance defined in Equation~\eqref{hasselmann}. 
Thus, for each, \( 1 \leqslant i \leqslant p \) and \( 1 \leqslant j \leqslant m \), we consider the associations: \( A_{i} \leftrightarrow \left(h(A_i), t(A_i), \theta (A_i), \sigma (A_i)\right) \) and \( B_{j} \leftrightarrow \left(h (B_j), t (B_j), \theta (B_j), \sigma (B_j)\right) \), to set our spectral distance (which is a weighted semimetric) \( d \) between the partitions \( A_i \) and \( B_j \) as
%Thus, for a generic partition \( \Lambda \), whose wave height is \( h(\Lambda) \), wave period is \( t(\Lambda) \), wave direction is \( \theta(\Lambda) \), and directional spreading is \( \sigma(\Lambda) \), will be identified with its wave parameter vector, that is: \(\Lambda \leftrightarrow (h(\Lambda), t(\Lambda), \theta(\Lambda), \sigma(\Lambda)).\)
%Bearing in mind the partitions of the measured wavenumber spectra \( A \) and \( B \), it is assumed, for notational simplicity, that for each \( 1 \leqslant i \leqslant p \) and each \( 1 \leqslant j \leqslant m \), we have \( A_{i} \equiv \left(h(A_i), t(A_i), \theta (A_i), \sigma (A_i)\right) \) and \( B_{j} \equiv \left(h (B_j), t (B_j), \theta (B_j), \sigma (B_j)\right) \). With this, we define the weighted semimetric \( d \) between the partitions \( A_i \) and \( B_j \) as:
\begin{equation}\label{semi-dist-pond}
d\left(A_{i}, B_{j}\right) =  \mathbf{v}\left(A_{i}, B_{j}\right) \cdot \mathbf{w}, 
\end{equation} 
%a dot product, where \(\mathbf{v}\left(A_{i}, B_{j}\right)\) is the variability vector, which carries the partitioned integrated wave parameters variabilities of \(A_i\) and \(B_j\), while \(\mathbf{w}\) is the weighting vector, whose entries assign weights to each partitioned integrated wave parameters. They are defined as follows:
a dot product, where \(\mathbf{v}\left(A_{i}, B_{j}\right)\) represents the variability vector, encapsulating the variabilities of the partitioned integrated wave parameters between \(A_i\) and \(B_j\), and \(\mathbf{w}\) is the weighting vector, whose components assign weights to each partitioned integrated wave parameter. These vectors are defined as:
\begin{equation}
    \mathbf{v}\left(A_{i}, B_{j}\right) = \left(v_h (A_i,B_j); \, v_t(A_i,B_j); \,v_\theta (A_i,(B_j); \, v_\sigma (A_i,B_j)\right),
\end{equation} where 
%\begin{equation}\label{semi-dist-pond}
%d\left(A_{i}, B_{j}\right) = w_h\, v_h(A_i,B_j) + w_t\, v_t(A_i,B_j)  + w_\theta \,\hat{v}_{\theta}(A_i,B_j) + w_\sigma v_{\sigma}(A_i,B_j)
%\end{equation}
%Here, the scalars \( w_h, \,w_t, \,w_\theta, \, w_\sigma >0 \) are (weights) such that \( w_h + w_t + w_\theta + w_\sigma = 1 \). The expressions \(v_z(A_i,B_j) = \nu [z(A_i),z(B_j)]\) for \(z=h, t\), and \(\sigma\) and \(\hat{\nu}_{\theta}(A_i,B_j) = \hat{v}[\theta(A_i),\theta(B_j)]\) are the variabilities associated with each pair of parameters combined in the matching process, and are defined through the relative variations \(\nu\) and \(\hat{\nu}\), given by:
\begin{equation}\label{variabildade}
    v_z[x, y]=\displaystyle\frac{|z(x)-z(y)|}{\max \{z(x), z(y)\}} \quad  \quad \text{for} \quad \quad z = h, t, \sigma
\end{equation} and 
\begin{equation}
  v_\theta[x,y]=\displaystyle\frac{1}{180^{\circ}} \min \{|\theta(x)-\theta(y)|, 360^{\circ} -|\theta(x)-\theta(y)|\}  
\end{equation} 
%the variation functions, both bounded above by \(1\). The weighting vector \(\mathbf{w} = (w_h,w_t,w_{\theta},w_{\sigma})\) is such its entries \(w_h,w_t,w_{\theta}\) and \(w_{\sigma}\) are positive scalars and satisfy \(w_h + w_t +w_{\theta} + w_{\sigma}= 1\). It is important to note that there are endless possibilities for the weighting vector: the reliability of the analyzed data, for example, is a rational criterion for defining each partitioned integrated wave parameter weight. Equation \eqref{semi-dist-pond} is explicitly the weighted sum of the partitioned wave parameters variabilities of \(A_i\) and \(B_j\), it reads,
are the variation functions, both bounded above by~1. The weighting vector \(\mathbf{w} = (w_h,w_t,w_{\theta},w_{\sigma})\) consists of positive scalars \(w_h,w_t,w_{\theta}\) and \(w_{\sigma}\), which satisfy the condition \(w_h + w_t +w_{\theta} + w_{\sigma}= 1\). 
It is important to emphasize that the weighting vector offers endless possibilities; for instance, the reliability of the analyzed data serves as a rational criterion for assigning weights to each partitioned integrated wave parameter.

Equation~\eqref{semi-dist-pond} explicitly represents the weighted sum of the partitioned wave parameter variabilities between \(A_i\) and \(B_j\), and is given as:
\begin{equation}\label{dist-explicita}
   d\left(A_{i}, B_{j}\right) = v_h\left(A_{i}, B_{j}\right) \, w_h +  v_t\left(A_{i}, B_{j}\right) \, w_t  + v_\theta\left(A_{i}, B_{j}\right)\, w_\theta + v_\sigma\left(A_{i}, B_{j}\right) \,w_\sigma. 
\end{equation} 
%If the weighting vector \(\mathbf{w}\) is balanced, that is, \(w_h = w_t = w_\theta = w_\sigma\), then the distance \eqref{dist-explicita} is reduced to the arithmetic mean of the variabilities of the partitioned wave parameters. Now, we observe that in contrast to the distance defined by the Equation \eqref{hasselmann}, if we have that \( d\left(A_{i}, B_{j}\right) = 0\), then \( A_{i} \) and \( B_{j} \) concord in all their wave parameters (of the same nature) since \(d\) is a semimetric; if \( d\left(A_{i}, B_{j}\right) \approx 1\) then at least one wave parameter (of the same nature) from \( A_{i} \) e \( B_{j} \) has a high discrepancy. 
If the weighting vector \(\mathbf{w}\) is balanced, meaning \(w_h = w_t = w_\theta = w_\sigma\), the distance in Equation~\eqref{dist-explicita} simplifies to the arithmetic mean of the variabilities of the partitioned wave parameters. 
Notably, in contrast to the distance defined by Equation~\eqref{hasselmann}, if \( d\left(A_{i}, B_{j}\right) = 0\), then \( A_{i} \) and \( B_{j} \) fully agree in all their corresponding wave parameters, as \(d\) is a semimetric. Conversely, if \( d\left(A_{i}, B_{j}\right) \approx 1\), it indicates that at least one corresponding wave parameter between \( A_{i} \) e \( B_{j} \) exhibits a high degree of discrepancy.

\subsubsection{A New Cross-Assignment Formulation}
% In this section, a new conception of cross-assignment is introduced. In effect, it will be shown how the algebraic structure of a weighted semimetric can be used to control the variability of the cross-assigned partitioned integrated wave parameters in a matchup produced from the cross-assignment process. 
% For this purpose, consider a control vector \(\mathbf{c}=(c_h,c_t,c_{\theta},c_{\sigma})\) whose coordinates satisfy \(0 \leqslant c_h,\, c_t,\, c_{\theta},\,c_{\sigma} \leqslant 1\). The \(n\)-matching  given by \eqref{matching} is the \textit{controlled four-parameter cross-assignment} between \(A\) and \(B\) relative to weighted semimetric \(d\) and subject to the control vector \(\mathbf{c}\) if the following conditions:

In this section, a novel framework for cross-assignment is introduced. Specifically, we demonstrate how the algebraic structure of a weighted semimetric can be employed to regulate the variability of the cross-assigned partitioned integrated wave parameters within a matchup generated by the cross-assignment process.
To achieve this, consider a control vector \(\mathbf{c}=(c_h,c_t,c_{\theta},c_{\sigma})\), where each coordinate satisfies \(0 \leqslant c_h,\, c_t,\, c_{\theta},\,c_{\sigma} \leqslant 1\). The \(n\)-matching defined by \eqref{matching} is referred to as the \textit{controlled four-parameter cross-assignment} between \(A\) and \(B\), relative to the weighted semimetric $d$ and governed by the control vector $c$, if the following conditions are satisfied:
%Consider a control vector, that is, any four-dimensional vector \(\mathbf{c}=(c_h,c_t,c_{\theta},c_{\sigma})\) whose coordinates satisfy \(0 \leqslant c_h,\, c_t,\, c_{\theta},\,c_{\sigma} \leqslant 1\). A \textit{four-controlled cross-assignment} between \(A\) and \(B\) induced by the weighted semimetric \(d\) and subject to the control vector \(\mathbf{c}\), when it exists, is a \(n\)-matching
%\begin{equation}\label{cross-assignment-controlado}
%\mathcal{P}_\mathbf{c} = \left\{\left(A_{k_1}, B_{j_{k_1}}\right) ;\left(A_{k_2}, B_{j_{k_2}}\right) ; \ldots ;\left(A_{k_n}, B_{j_{k_n}}\right)\right\},
%\end{equation} with \(n \leqslant p\) the largest possible and such that the following conditions:
%\begin{enumerate}
%    \item[(c1)] if \(1\leqslant i \neq i'\leqslant n\) then \(1\leqslant k_i\neq k_{i'}\leqslant p\) and \( 1 \leqslant j_{k_{i}} \neq j_{k_{i'}} \leqslant m\).\vskip0.5cm  
 %   \item[(c2)] if \(1 \leqslant i \leqslant n\) then \(\mathbf{v}(A_{k_i},B_{j_{k_i}})\leqslant \mathbf{c}\). \vskip0.5cm  
  %  \item[(c3)] \(\displaystyle\sum_{i=1}^{n} d\left(A_{k_i}, B_{j_{k_i}}\right) = \min \left\{ \displaystyle \sum_{i=1}^{n} d\left(A_{\rho(i)}, B_{\tau(\rho(i))}\right): \rho \in I_{n,p}, ~ \tau \in I_{p, m} \right\}.\)
%\end{enumerate}
\begin{enumerate}
    \item[(i')] \(n \leqslant p\) is the largest possible
    \item[(ii')] \(\mathbf{v}(A_{i_k},B_{j_k})\preceq \mathbf{c}\)  \footnote{the symbol \(\preceq\) indicates that the components of the variability vector do not exceed the corresponding components of the control vector.} for each \(i=1,2,\ldots,n\) \vskip0.5cm  
    \item[(iii')] \(\displaystyle\sum_{k=1}^{n} d\left(A_{i_k}, B_{j_k}\right) = \min \left\{ \displaystyle \sum_{k=1}^{n} d\left(A_{\rho(k)}, B_{\tau(\rho(k))}\right): \rho \in I_{n,p}, ~ \tau \in I_{p, m} \right\}\)
\end{enumerate} are valid\footnote{the symbols \(I_{n,p}\) and \(I_{p,m}\) represent, respectively, the sets of all injective functions from \(\{ 1,2,\ldots,n \}\) into \(\{ 1,2,\ldots, p\}\) and from \(\{ 1,2,\ldots, p\}\) into \(\{ 1,2,\ldots, m\}\).}.

%Condition (c1) ensures that the set of pairs seen in \eqref{cross-assignment-controlado} is a matching, that is, each partition of \(A\) is matched to at most one partition of \(B\). Condition (c2) provides an accuracy control for the cross-assigned partitioned wave parameters of the matchups; in this sense, note that the previously chosen constraining scalars \(c_h,c_t,c_{\theta}\), and \(c_{\sigma}\) play a fundamental role in this context since ultimately they control the distance between the matched partitions. In fact, if the partitions \(A_{k_i}\) and \(B_{j_{k_i}}\) are \(\mathbf{c}\)-controlled ~---~i.e., condition (c2) is valid ~---~ then the distance between these partitions is such that

Condition (ii') provides control for the cross-assigned partitioned wave parameters of the matchups; the previously chosen constraining scalars \(c_h,c_t,c_{\theta}\), and \(c_{\sigma}\) control the distance between the matched partitions. 
If the partitions \(A_{i_k}\) and \(B_{j_{k}}\) are \(\mathbf{c}\)-controlled~---~i.e., condition (ii') is valid~---~then the distance between these partitions is
\begin{equation}
d \left(A_{i_k},B_{j_k}\right) \leqslant w_h c_h + w_t c_t + w_{\theta} c_{\theta} + w_{\sigma} c_{\sigma}    
\end{equation} 
The optimality condition (iii') characterizes controlled cross-assignment as the matching between \(A\) and \(B\) with the shortest possible length.
The quantity calculated in (iii') represents the length of the cross-assignment; among all the \(n\)-matchings of \(\mathbf{c}\)-controlled partitions of \(A\) and \(B\), this is the one with the smallest length.

%Regarding the existence, we observe that if \(c_h = c_t = c_{\theta} = c_{\sigma} = 1\), the cross-assignment problem is obviously unconstrained, and in this case, there is always a solution. On the other hand, for example, if \(c_h = c_{\theta} = c_{\sigma} = 1\) and \(c_t = 0.3\), there is no constraint on the values of the significant wave heights, nor on the values of the peak wave directions, nor the values of the peak directional spreading of the matched partitions, but there is a relevant constraint on the values of the peak wave periods of the matched partitions~---~the period of one of the partitions is at least\footnote{it comes from Equation \eqref{variabildade}} \(70\%\) of the period of the other partition. In this case, the existence of a constrained cross-assignment depends on the dataset being analyzed. Therefore, an appropriate adjustment of the control vector may be essential to obtain solutions to a controlled cross-assignment problem for a particular data set.

If \(c_h = c_t = c_{\theta} = c_{\sigma} = 1\), the cross-assignment problem is unconstrained, and in this case, there is always a solution. 
%On the other hand, for example, if \(c_h = c_{\theta} = c_{\sigma} = 1\) and \(c_t = 0.3\), there is no constraint on the values of the significant wave heights, nor on the values of the peak wave directions, nor the values of the peak directional spreading of the matched partitions, but there is a relevant constraint on the values of the peak wave periods of the matched partitions.
Conversely, for instance, if \(c_h = c_{\theta} = c_{\sigma} = 1\) and \(c_t = 0.3\), there are no constraints on the values of significant wave heights, peak wave directions, or peak directional spreading for the matched partitions. However, a significant constraint is imposed on the values of peak wave periods for the matched partitions.
In each matchup, the smallest period is at least $(1 - 0.3)=0.7$ of its cross-assigned counterpart.
The largest period of that matchup~---~or in other words, the smallest period is at least 70\% of the largest period. In this case, the existence of a constrained cross-assignment depends on the dataset being analyzed. Therefore, an appropriate adjustment of the control vector may be essential to obtain solutions to a controlled cross-assignment problem for a particular data set.

%We conclude this section by summarizing that solving a four-controlled cross-assignment problem means finding a cross-assignment between two measurements of the wave spectrum of a given sea state, in which the values of partitioned wave parameters of connected wave systems have a priori controlled variation.

%In summary, the \textit{four-controlled cross-assignment} problem consists of: given two partitioned wave spectra (measured) of a particular sea state, find the cross-assignment between its partitions induced by a weighted semimetric and such the four partitioned wave parameters of linked partitions are - a priori - controlled by means of four control scalars\footnote{i.e. the control vector}. We developed a computational routine to solve this problem is the so-called \textit{controlled four-parameter method} for cross-assignment.

% In summary, in this section, we defined a conception of cross-assignment, which essentially consists of: given two partitioned wave spectra (measured) of a particular sea state, finding the \textit{best} matching between their spectral partitions in which all formed matchups have their four partitioned integrated wave parameters discrepancies controlled a priori. To solve this problem, we developed a computational routine called \textit{controlled four-parameter method} (abbreviated, C4PM), whose diagram is shown in Figure \ref{fig:C4PMdiag}. In particular, write u-C4PM to indicate that C4PM is operating in uniform mode, i.e. the control vector is uniform, which means \(c_h = c_t = c_{\theta} = c_{\sigma} = r\). 

\begin{figure}[ht!]    
    \centering
    \begin{subfloat}{
        \centering
        \includegraphics[width=13cm]{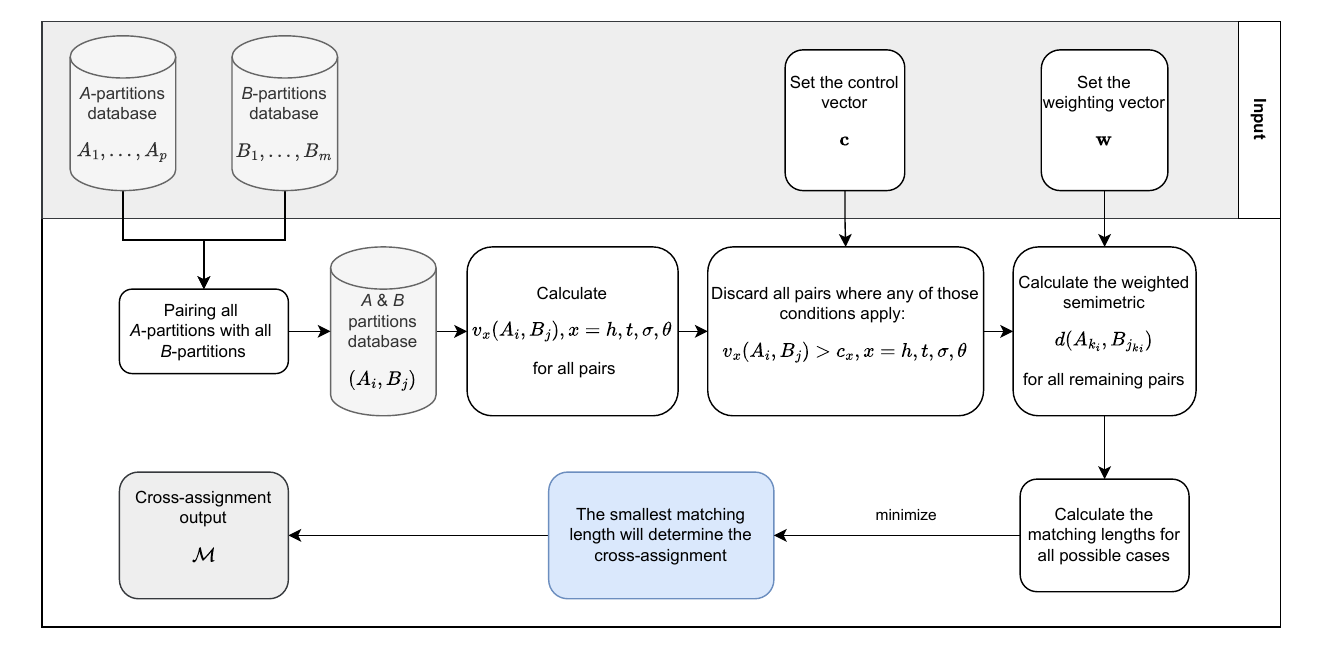}}
    \end{subfloat}
    \caption{C4PM algorithm diagram.}\label{fig:C4PMdiag}
\end{figure}

In summary, this section introduced a framework for cross-assignment, which consists of the following: given two partitioned wave spectra (measured) of a specific sea state, the objective is to find the \textit{best} matching between their spectral partitions, ensuring that the discrepancies in all four partitioned integrated wave parameters are controlled \textit{a priori}. To address this problem, we developed a computational routine called the \textit{Controlled Four-Parameter Method} (C4PM), whose diagram is presented in Figure~\ref{fig:C4PMdiag}. Specifically, we denote the method as u-C4PM when it operates in uniform mode, meaning the control vector is uniform, with \(c_h = c_t = c_{\theta} = c_{\sigma} = r\).

%Finally, it is noted that the cross-assignment resulting from C4PM is not necessarily equal to that resulting from 2PM, mainly due to the way in which its partition pairs are formed, as well as the spectral distance peculiar to each method.

\section{Data and Methods} \label{MET}

%This section describes the dataset from the wave buoys and the proposed methodology for comparing both cross-assignment techniques.

\subsection{NDBC Buoy Data}

\begin{figure}[ht!]    
    \centering
    \begin{subfloat}{
        \centering
        \includegraphics[width=10.5cm]{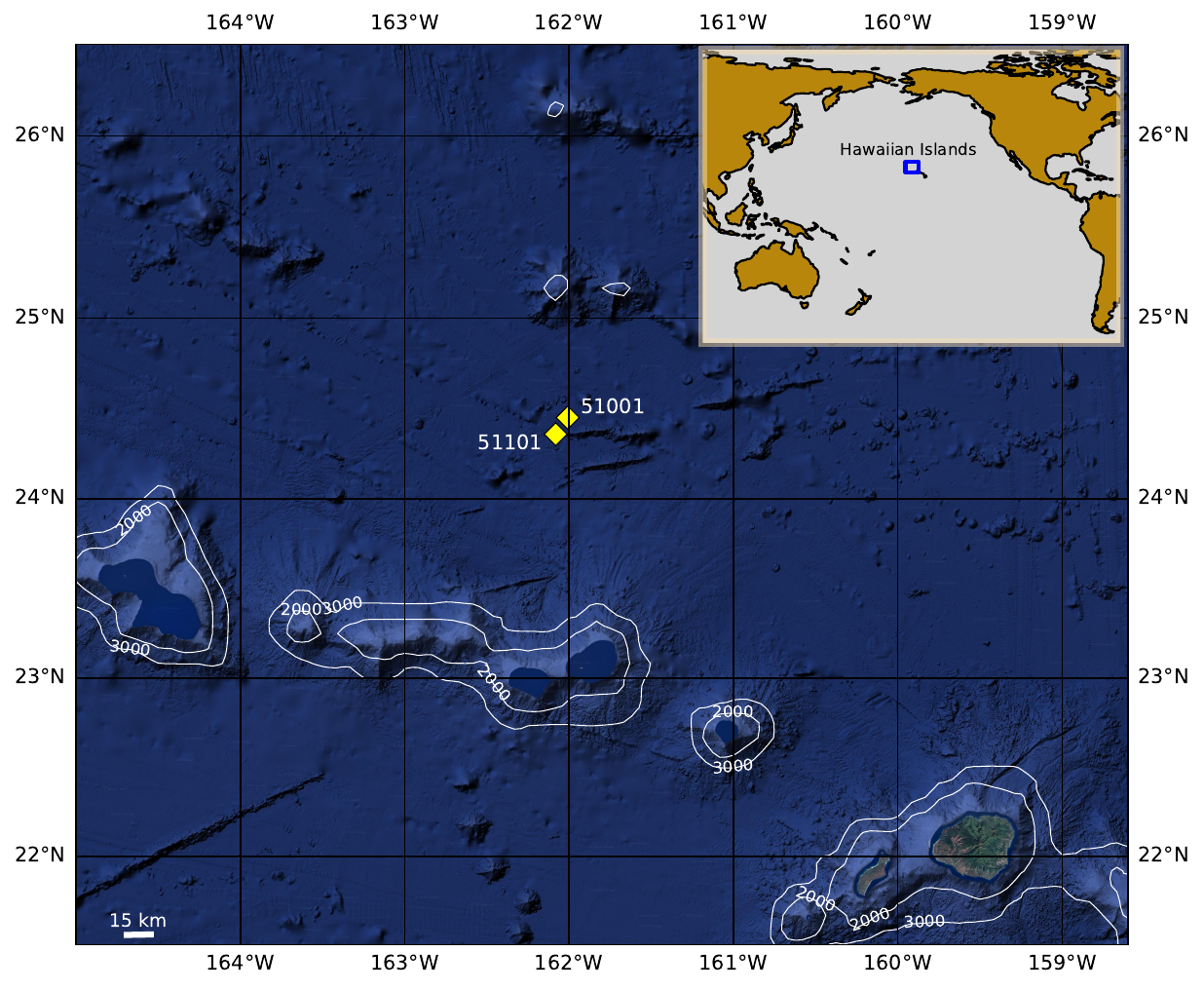}}
    \end{subfloat}
    \caption{Location of the NDBC buoys 51001 and 51101, which are 13~km apart, in deep water off Hawaii.}\label{fig:map_buoys2}
\end{figure}

We selected two buoys, operated by the National Data Buoy Center (NDBC), with IDs 51001 and 51101, for the period from January 2023 to December 2023~---~details available at https://www.ndbc.noaa.gov/.
These buoys are situated approximately 13~km apart in deep water off the coast of Hawaii (Figure~\ref{fig:map_buoys2}). 
Their proximity and location in similar deep-water conditions indicate that both buoys provide relatively comparable measurements. This implies a reasonable number of high-quality pairs can be identified between them, while also reducing challenges associated with applying cross-assignment methodologies, as will be discussed in the following sections.
Both buoys collect meteorological and oceanographic data, including the directional wave spectra necessary for cross-assignment techniques. 
Wave data from these buoys are available at 30~minute intervals, yielding a total of 17,100 pairs of wave spectra (Table~\ref{tab:tab_buoys}). % provides basic information about the buoys.

\begin{table}[ht!]
  \begin{center}
    \label{tab:tab_buoys}
    %\begin{tabular}{cccc} % <-- Changed to S here.
    \begin{tabular}{|c c c c| }
     %\toprule
      \hline
      ID & Latitude & Longitude & Depth\\     
      \hline
      %\midrule
     51001   & 24.451~N & 162.008~W & 4906~m \\
     51101   & 24.359~N & 162.081~W & 4860~m \\
     \hline
      %\bottomrule
      \end{tabular}
      \caption[Buoys Information]{NCBC Buoys.}
  \end{center}
\end{table}

\begin{figure}[ht!]
    \centering
    \begin{subfloat}{
        \centering
        \includegraphics[width=6cm]{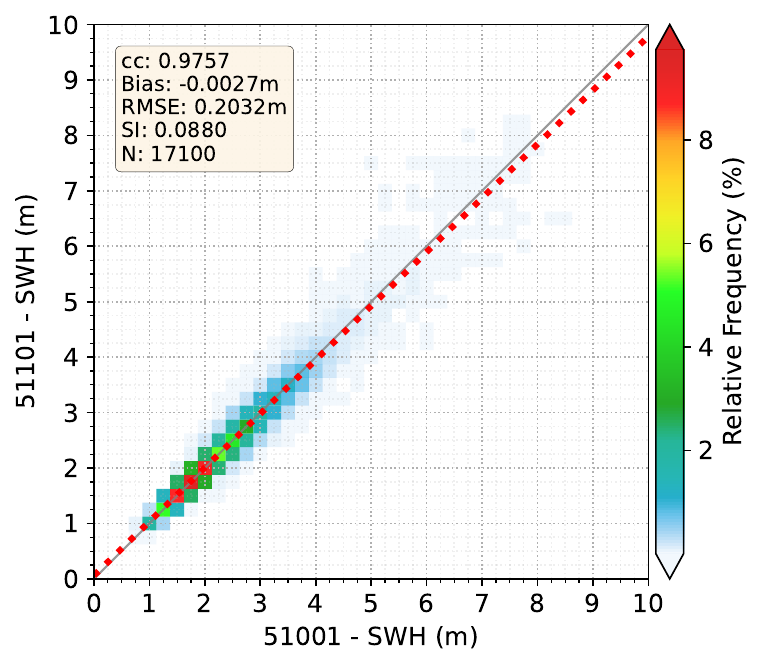}
        \label{subfig:buoy_scat_hs}}
        \caption{Significant Wave Height ($SHW$)}
    \end{subfloat}
    %\hfill
    \hspace*{0.5cm}
    \begin{subfloat}{
        \centering
        \includegraphics[width=6cm]{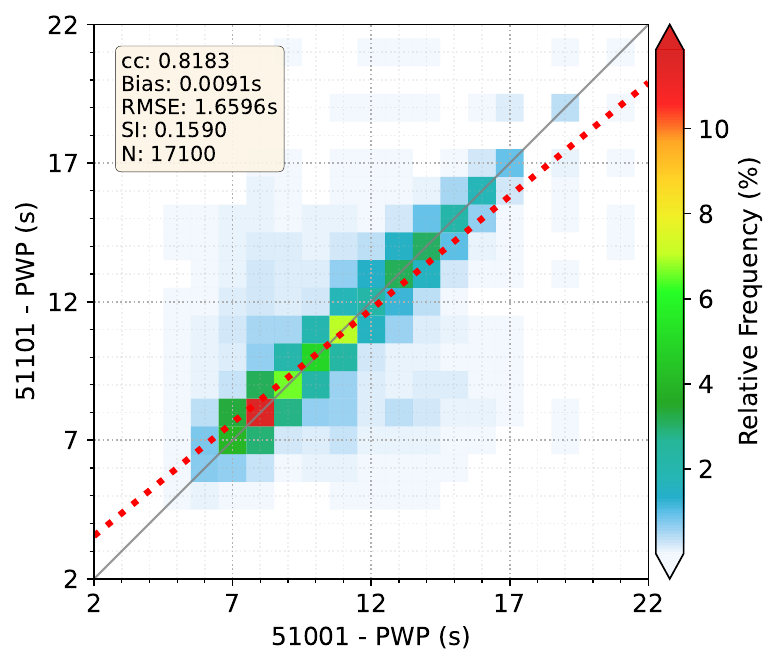}
        \label{subfig:buoy_scat_tp}}
        \caption{Peak Wave Period ($PWP$)}
    \end{subfloat}
    
    \begin{subfloat}{
        \centering
        \includegraphics[width=6cm]{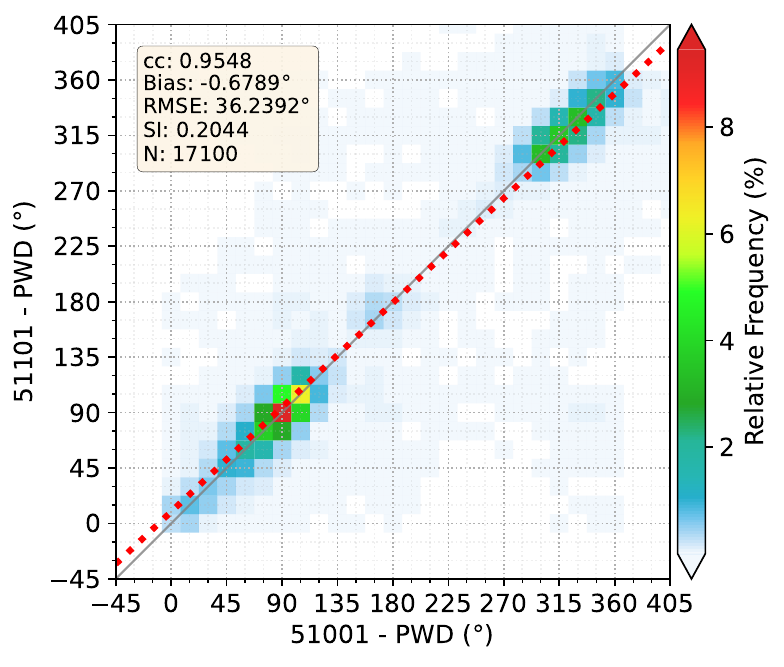}
        \label{subfig:buoy_scat_dp}}
        \caption{Peak Wave Direction ($PWD$)}
    \end{subfloat}
    %\hfill
    \hspace*{0.5cm}
    \begin{subfloat}{
        \centering
        \includegraphics[width=6cm]{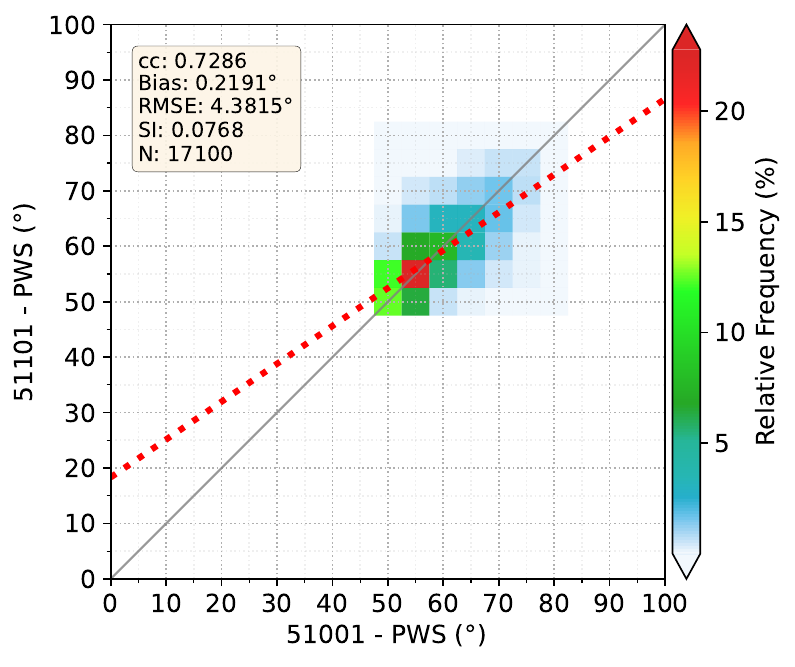}
        \label{subfig:buoy_scat_spr}}
        \caption{Peak Directional Spreading ($PWS$)}
    \end{subfloat}
\caption{Scatter plots between buoys 51001 and 51101 over the year 2023. (a) $SWH$, ~(b) $PWP$,~ (c) $PWD$ and (d) $PWS$. The boxes indicate Pearson correlation coefficient ($cc$), Bias, Root Mean Square Error ($RMSE$), Scatter Index ($SI$) and total number of points ($N$).}
\label{fig:buoy_scat}
\end{figure}

Figure~\ref{fig:buoy_scat} shows, over the selected time period of 2023, the scatter plots of wave parameters: significant wave height $(SWH)$, peak wave period $(PWP)$, peak wave direction $(PWD)$ and peak wave spreading $(PWS)$. %, and only the most energetic system is depicted.
The wave parameters were downloaded directly from the NDBC page~---~with exception of $PWS$ which was computed as outlined in Table~\ref{tab:table2}~---~therefore the spectra were not partitioned. The statistical parameters are listed in Table~\ref{tab:stats}.

%In a sense, this could be perceived as a cross-assignment based on the energy ranking of the most energetic partition.
As expected, the biases across the four parameters are minimal. 
However, low-frequency occurrences representing significant differences among the spectral peak pairs are evident, as indicated by the light blue shades, particularly in relation to direction and period (Figure~\ref{fig:buoy_scat}b~and~c).
For $SWH$, the $RMSE$ is 0.20~m with a correlation coefficient equal to $0.98$, indicating a strong level of similarity. 
Although the agreement for $PWP$ is generally good, numerous outliers in Figure~\ref{fig:buoy_scat}b contribute to an increased $RMSE$ and lower correlation coefficient.
A similar pattern is seen in Figure~\ref{fig:buoy_scat}c, where the outliers introduce substantial discrepancies.
The lower correlation coefficient among the four wave parameters is associated with $PWS$, a parameter that is challenging to estimate accurately from single point measurements such as directional buoys \citep{kuik}.

\subsection{Data Processing}

The NDBC provides wave spectral data in the form of five parameters concerning the Fourier series expansion of the buoy's directional wave spectrum: the non-directional spectral density, the first normalized directional Fourier coefficient, the mean wave direction, the second normalized directional Fourier coefficient and the main wave direction. The time series for each buoy were downloaded for the year~2023.
The data are then time-synchronized and consolidated into a database, removing any incomplete entries (i.e. missing data) to ensure that each record has complete information for all five wave parameters from both buoys, at the same date and time.
Each row of this database is then processed following \cite{Earle1999UseOA}, allowing the calculation of the weighted directional spreading function and directional wave spectrum for each buoy. The weighted directional spreading function ensures that the calculated directional wave spectra remain non-negative, addressing potential problems with negative values that can arise when using more basic approaches.

%The NDBC provides wave spectral data in the form of five parameters derived from the Fourier series expansion of the buoy’s directional wave spectrum, following \cite{longuet1963observation}. 
%These parameters are: non-directional spectral density, first normalized directional Fourier coefficient, mean wave direction, second normalized directional Fourier coefficient, and principal wave direction. 
%The first step in the analysis was to download the time series of these parameters for each buoy for the year 2023 from the NDBC website.

%Next, the data were time-synchronized and consolidated into a database, with any incomplete entries (i.e., missing data) removed to ensure that each record contained complete information for all five parameters from both buoys at the same date and time. 
%Each row of this database was then processed following the methodology of \cite{Earle1999UseOA}, allowing for the calculation of the weighted directional spreading function and the directional wave spectrum for each buoy. 
%Notably, the use of the weighted directional spreading function ensures that the calculated directional wave spectra remain non-negative, addressing potential issues with negative values that can arise when using more basic approaches.

%Once the directional wave spectra were available, they were partitioned, all the partition wave parameters were calculated ($H_{s}$, $T_{p}$, $\phi_{p}$, $spr$) according to Table \ref{tab:table2}, and the noisy partitions were removed. This last process considered two criteria to classify a partition as noisy:

Once the directional wave spectra is obtained, all partitioned integrated wave parameters are calculated as outlined in Table~\ref{tab:table2}, and all noisy partitions are removed according to the following criteria:

%\begin{enumerate}
%    \item[(a)] Partition significant wave height less than 0.25m;
    
%    \vskip0.5cm  

%    \item[(b)] Partition peak period less than 5 seconds and partition significant wave height less than 10\% of the full-spectrum $H_s$.
%\end{enumerate}

\begin{enumerate}
    \item[(a)] $PSWH \leqslant 0.25$\,m;
    
    \vskip0.5cm  

    \item[(b)] $PPWP \leqslant 5$\,s and $PSWH \leqslant 10$\% $SWH$.
\end{enumerate}

\begin{figure}[ht!]    
    \centering
    \begin{subfloat}{
        \centering
        \includegraphics[width=13cm]{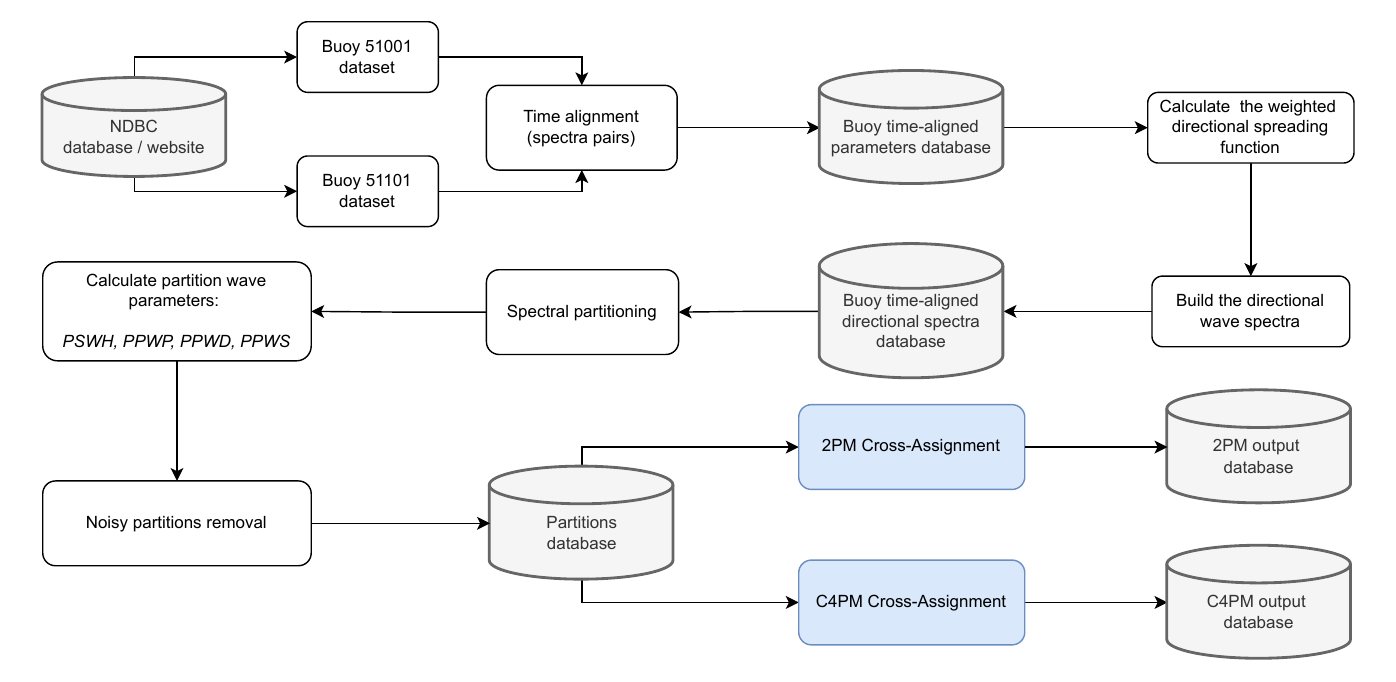}}
    \end{subfloat}
    \caption{Data processing diagram}\label{fig:data_diagram}
\end{figure}

At this stage, the partition database is consolidated, with the greatest possible amount of matches being 30,956. 
Both cross-assignment techniques, 2PM and C4PM, were applied to the same dataset, each generating its respective output databases, as illustrated in the processing workflow shown in Figure~\ref{fig:data_diagram}.
%The data analysis focuses primarily on the percentage of cross-assigned partitions, along with various statistical indicators for all wave parameters (as detailed in Table~\ref{tab:stats}), and includes scatter plots comparing the results between both buoys and cross-assignment methods.

\section{Results} \label{res}

%In general, cross-assignment tasks employ spectra from different sources, such as numerical models, \textit{in-situ} or remote sensing measurements~---~which, naturally, have some degree of discrepancy among them.
%In many cases, the number of partitions in each paired spectrum differs, with some partitions missing and others being spurious.
%In these common cases, the cross-assignment has to rely heavily on the capability of the method employed.

In general, cross-assignment tasks employ spectra from different sources, such as numerical models, \textit{in situ} measurements, or remote sensing, which naturally have some degree of discrepancy between them. In many cases, the number of partitions in each paired spectrum differs, with some partitions missing and others being spurious. In these common cases, cross-assignment has to rely heavily on the ability of the employed method to distinguish good matchups. In this sense, some experiments are proposed and analyzed to demonstrate the abilities of C4PM, taking 2PM as a reference. 

It is important to note that all C4PM experiments described in this section rely on a balanced weight vector to establish a general approach for comparison with the reference method.

%\subsection{Cross-assignment evaluation}
%It's a central issue to validate C4PM and evaluate its performance. To this end, we draw attention to the parallelism between the roles played by the critical value \(R\) with respect to 2PM and the uniform control value \(r\) with respect to u-C4PM: ultimately, both are cutoff values and, in absolutely different ways, control the distance between partitions that have been matched in each of the cross-assignment processes. %So, in this section, two experiments are proposed to establish a first comparative analysis between 2PM and C4PM.

\subsection{Sensitivity test}
%It's a central issue to validate C4PM and evaluate its performance. To this end, we draw attention to the parallelism between the roles played by the critical value with respect to 2PM and the control vector with respect to C4PM: ultimately, both are cutoff values and, in absolutely different ways, control the distance between partitions that have been matched in each of the cross-assignment processes. 
\begin{table}[ht!]
    \centering
    \begin{tabular}{|c c c c c | }
    \hline
    \multicolumn{5}{|c|} {\small{2PM: matchups distribution}} \\
    \cline{1-5}
    
          \small{$l$} & {} & \small{$R_l$} & {} & \small{$Q_l$} \\ 
          \hline
           1 & {} & 0.2 & {} & 26213 \\
           2 & {} & 0.4 & {} & 28321 \\
           3 & {} & 0.6 & {} & 29192 \\
           4 & {} & 0.8 & {} & 29654 \\
           5 & {} & 1.0 & {} & 30103 \\
           6 & {} & 1.2 & {} & 30454\\
           7 & {} & 1.4 & {} & 30649\\
           8 & {} & 1.6 & {} & 30744\\
           9 & {} & 1.8 & {} & 30850\\
           10& {} & 2.0 & {} & 30956\\
          \hline
    \end{tabular}
    \caption{2PM progression of the number of matchups against equidistant critical values.}% for a maximum possible number of matchups $N_{\max} = 30,956$.}
    \label{2PM-distribution}
\end{table} 

% It's a central issue to validate C4PM and evaluate its performance. To this end, we draw attention to the parallelism between the roles played by the critical value (\(R\)) with respect to 2PM and the control vector (\(\mathbf{c}\)) with respect to C4PM: ultimately, both are cutoff quantities and, in absolutely different ways, control the distance between partitions that have been matched in each of the cross-assignment processes. 
% Validating C4PM and evaluating its performance is a central issue. To this end, we highlight the parallelism between the roles of the critical value (\(R\)) in 2PM and the control vector (\(\mathbf{c}\)) in C4PM. Ultimately, both serve as cutoff parameters and, albeit through fundamentally different mechanisms, regulate the distance between partitions matched during each cross-assignment process.
% In this test, the distributions of the matchups emanating from u-C4PM and 2PM algorithms with equally spaced consecutive cutoff values are presented. This will allow us to understand the sensitivity of each of the spectral distances operating behind these methods. 
% For this purpose, we define the sequence of cutoff values given by:
Validating C4PM and assessing its performance is a central issue. To this end, we highlight the parallelism between the roles of the critical value (\(R\)) in 2PM and the control vector (\(\mathbf{c}\)) in C4PM. Ultimately, both act as cutoff parameters and, albeit through fundamentally different mechanisms, regulate the distance between partitions matched in each cross-assignment process.

We define the sequence of cutoff values as follows: 
\begin{equation}
\left\{\begin{array}{lllllll}
  R_l & = & \frac{l}{10} & R_{\max} &  & \text{(critical value)} & \\
  
  r_l & = & \frac{l}{10} & r_{\max} &  & \text{(control value)} & 
\end{array}\right.,   
\end{equation} for $l=1,2,\ldots,10$. 
The $l$-th experiment in this test involves determining the number of matchups $Q_l$ and $q_l$ generated by the 2PM and u-C4PM\footnote{In uniform mode, the control vector is of the form $\mathbf{c} = r(1,1,1,1)$.} runs, respectively, when initialized with input data \( R_l\) and \( r_l\). 
It can be observed that $R_{\max} = 2$ (for the selected dataset) and $r_{\max} = 1$ correspond, in this order, to the maximum possible number of matchups $Q_{\max}$ and $q_{\max}$, both of which are equal to $N_{\max} = 30,956$ (reached for $l=10$).
% The results of this test are shown in Tables \ref{2PM-distribution} and \ref{C4PM-distribution}, which immediately reveal an important distinction between the spectral distance behaviors of C4PM and 2PM. In fact, regarding the distribution of matchups performed by C4PM, Table \ref{C4PM-distribution} initially presents a progressive and scaled distribution of matchups from the 1st to the 5th metric deciles; after that, a certain stabilization is observed until the 10th metric decile. In contrast, the distribition of matchups performed by 2PM, according to Table \ref{2PM-distribution}, starts with a very high value in the 1st experiment and the jump until the 10th metric decile is relatively small. Add to this that the amounts of matchups corresponding to the 1st metric decile of 2PM and the 4th metric decile of C4PM are quite close. 
% Taking these data together, this suggests that weighted semimetric, the basis of C4PM, have a better ability to distinguish matches. This fact will become even clearer in the following sections.
The results are presented in Tables~\ref{2PM-distribution} and \ref{C4PM-distribution}, which highlight a key distinction in the spectral distance behaviors of C4PM and 2PM. 
Specifically, the distribution of matchups performed by C4PM, as shown in Table~\ref{C4PM-distribution}, exhibits a progressive and scaled increase from the 1st to the 5th metric deciles, followed by a stabilization trend up to the 10th metric decile. 
In contrast, the distribution of matchups performed by 2PM, according to Table~\ref{2PM-distribution}, begins with a very high value in the 1st metric decile, with only a relatively small increase observed up to the 10th metric decile.
Moreover, the number of matchups in the 1st metric decile of 2PM is remarkably close to the number of matchups in the 4th metric decile of C4PM. Taken together, these observations suggest that the weighted semimetric underlying C4PM offers a superior capability to distinguish between matches. 
This will become even more apparent in the subsequent sections.

%It means that C4PM has a greater ability to detect ocean features between two wave systems; this statement is reinforced when checking the root mean square error ($RMSE$) of each partitioned integrated wave parameter in both methods: the matchups classes generated at each experiment by C4PM systematically present a lower $RMSE$ compared to the $RMSE$ of the matchups generated by 2PM. Now, observing the 11th experiment, in which there are no restrictions, both techniques had similar performance regarding $RMSE$ values, but with C4PM still showing a slight advantage.

%As a step towards assessing the sensitivity of both methods, let \(R_{\max}\) and \(r_{\max}\) be a critical value and a uniform control value that return the greatest possible amount of matchups using HM and u-C4PM respectively. Observe that since the weighted semimetric \(d\) is bounded above, it is automatic that can be chosen \(r_{\max} = 1\). On the other hand, the critical value \(R_{\max} = 2\) is reached (and chosen) experimentally after a few 2PM runs.

\begin{table}[ht!]
    \centering
    \begin{tabular}{|c c c c c | }
    \hline
    \multicolumn{5}{|c|} {u-C4PM: matchups distribution} \\
    \cline{1-5}
    
          \small{$l$} & {} & \small{$r_l$} & {} & \small{$q_l$} \\ 
          \hline
           1 & {} & 0.1 & {} & 7338 \\
           2 & {} & 0.2 & {} & 18430 \\
           3 & {} & 0.3 & {} & 24112 \\
           4 & {} & 0.4 & {} & 27236 \\
           5 & {} & 0.5 & {} & 29177 \\
           6 & {} & 0.6 & {} & 29915\\
           7 & {} & 0.7 & {} & 30463\\
           8 & {} & 0.8 & {} & 30799\\
           9 & {} & 0.9 & {} & 30919\\
           10& {} & 1.0 & {} & 30956\\
          \hline
    \end{tabular}
    \caption{u-C4PM progression of the number of matchups against equidistant control values.}% To illustrate, $r_l=0.1$ signifies that the distance between partitions does not exceed 10\% for all four parameters simultaneously.}
    \label{C4PM-distribution}
\end{table}

\subsection{Accuracy test}\label{fac} 

As a step toward evaluating the performance of C4PM, this test analyzes the results obtained by running the 2PM and u-C4PM algorithms, each initialized with input values that yield the top 0.3\% (equivalent to 99 pairs) of the best matchups generated by each technique (Table~\ref{case-99}).
The input cutoff values were determined and, as expected, are very small: $R = 4.7 \times 10^{-5}$ (the critical value for 2PM) and $r = 1.0 \times 10^{-2}$ (the control value for u-C4PM).
Because the matchups generated by u-C4PM are $r$-uniformly controlled, it follows that, for all cross-assigned partitioned integrated wave parameters (except for partitioned wave peak directions), the smaller parameter is at least 99\% of the larger one. 
Additionally, the partitioned and cross-assigned wave peak directions differ by no more than \(1.8^{\circ}\). 
In contrast, among the 99 matchups produced by 2PM, 51 pairs exhibit at least one cross-assigned partitioned integrated wave parameter (excluding partitioned wave peak directions) where the smaller parameter is less than 90\% of the larger one. 
For partitioned wave peak directions, the difference in these cases is no less than \(18^{\circ}\).
These discrepancies are evidently reflected in the RMSEs of the partitioned integrated wave parameters, as summarized in Table~\ref{case-99}, and serve to justify the results presented therein.
Although the contrasted matchup groups lie within a particularly narrow range, this result highlights a key distinction in the performance of the methods: C4PM successfully prevented the formation of poorly matched partitions, whereas 2PM, even when operating at an extremely low critical value, did not.

\begin{table}[ht!]
    \centering
    \begin{tabular}{|r r r r r|}
     \hline
     \multicolumn{5}{|c|}{$RMSE$: a strict range}\\
     \cline{1-5}
        \small{Method} &  \small{$PSWH$ (m)} & \small{$PPWP$ (s)} & \small{$PPWD$ ($^{\circ}$)} & \small{$PPWS$ ($^{\circ}$)}\\ \hline 
        \small{2PM}     &   0.18  & 0.69 & 13.57 & 4.91\\
        \small{u-C4PM}   &   0.01  & 0.00 & 9.48 & 0.31\\
     \hline   
    \end{tabular}
    \caption{Contrasting u-C4PM and 2PM in refined settings.} %$RMSE$ of the 99 pairs out of $N_{\max}=30,956$ of the best matchups generated by each technique.} %The $RMSE$ employing C4PM approaches much smaller values compared to 2PM~---~see also Figure~\ref{fig:$RMSE$_N2}.}
    \label{case-99}
\end{table}

% Although the contrasted matchups groups are in a particularly narrow range, a first distinction between the performances of the methods is delimited from this result: C4PM prevented the formation of poorly matched partitions, while 2PM did not, even operating at an extremely low critical value.

\subsection{A broader comparison}
% In this test, the greatest amount of all possible matchups obtained by u-C4PM and 2PM runs will be strategically divided into quintiles and their $RMSEs$ will be displayed and discussed.
% Considering this, let $Q^{*}_l$ and $q^{*}_l$ be the amounts of matchups resulting from the runs of the 2PM and C4PM algorithms with input data $R^{*}_l$ and $r^{*}_l$ respectively. These cutoff values are defined by the equations

In this test, the largest set of all possible matchups obtained from the u-C4PM and 2PM runs will be divided into quintiles, and their corresponding $RMSEs$ will be presented and analyzed.
To formalize this, let $Q^{*}_l$ and $q^{*}_l$ represent the number of matchups produced by the 2PM and C4PM algorithms, respectively, when initialized with input data $R^{*}_l$ and $r^{*}_l$. These cutoff values are defined so that the equations
\begin{equation}
    Q^{*}_l = q^{*}_l = \frac{l}{5} N_{\max},  
\end{equation} for $l= 1, \ldots, 5$ are valid. 

Tables~\ref{2PM-per-quintil} and \ref{C4PM-per-quintil} summarize the performances of 2PM and C4PM across all quintiles. 
For both methods, the $RMSE$ values of all integrated wave parameters do not consistently decrease with each advancing quintile as the cutoff values progress. 
Notably, these tables partially reflect the results discussed in~\ref{fac}.
Consider the first quintile of each method. 
The 6191 matchups generated by C4PM are $r^{*}_1$-uniformly controlled with $r^{*}_1 = 0.0851$.
Therefore, for all pairs of cross-assigned integrated wave parameters, the smaller value is at least 91\% of the larger one, except for cross-assigned partitioned peak wave directions, whose differences do not exceed $15.4^{\circ}$. 
In contrast, these conditions do not hold for the 6191 matchups corresponding to the first quintile under 2PM, as evidenced by their higher $RMSE$ values. 
Specifically, 3647 matches include at least one cross-assigned partitioned integrated wave parameter where the smaller value is less than 91\% of the larger one, excluding wave peak directions, which differ by at least $15.4^{\circ}$.
This discrepancy accounts for the higher $RMSE$ values observed for 2PM compared to C4PM within this quintile.

\begin{table}[ht!]
    \centering
    \begin{tabular}{|c c c c c c c| }
    \hline
    \multicolumn{7}{|c|} {\small{2PM: $RMSE$ per quintile}} \\
    \cline{1-7}
    
          \small{$l$} &\small{$Q^{*}_l$} & \small{$R^{*}_l$} & \small{$PSWH$ (m)} & \small{$PPWP$ (s)} & \small{$PPWD$ ($^{\circ}$)} & \small{$PPWS$ ($^{\circ}$)} \\ 
          \hline
           1 & 20\% & 0,0044 & 0.22 & 0.72 & 13.17 & 4.59\\
           2& 40\% & 0,0142 & 0.23 & 0.76 & 14.81 & 5.00\\
           3& 60\% & 0,0415 & 0.26 & 0.81 & 17.11 & 5.63\\
           4& 80\% & 0,1420 & 0.29 & 0.92 & 20.24 & 6.81\\
           5& 100\% & 2,0000 & 0.32 & 1.80 & 34.74 & 9.22\\ 
          \hline
    \end{tabular}
    \caption{2PM performance for quintilian critical values.}
    \label{2PM-per-quintil}
\end{table}

\begin{table}[ht!]
    \centering
    \begin{tabular}{|c c c c c c c| }
    \hline
    \multicolumn{7}{|c|} {\small{u-C4PM: $RMSE$ per quintile}} \\
    \cline{1-7}
    
          \small{$l$} & \small{$q^{*}_l$} & \small{$r^{*}_l$} & \small{$PSWH$ (m)} & \small{$PPWP$ (s)} & \small{$PPWD$ ($^{\circ}$)} & \small{$PPWS$ ($^{\circ}$)} \\ 
          \hline
           1 & 20\% & 0.0851 & 0.11 & 0.00 & 9,48 & 2.23\\
           2 & 40\% & 0.1266 & 0.15 & 0.72 & 9,70 & 2.92\\
           3 & 60\% & 0.2050 & 0.20 & 0.74 & 13,28 & 4.05\\
           4 & 80\% & 0.3258 & 0.26 & 0.90 & 17,08 & 5.78\\
           5 & 100\% & 1.0000 & 0.31 & 1.65 & 34,71 & 8.50\\ 
          \hline
    \end{tabular}
    \caption{u-C4PM performance for quintilian (uniform) control values.} %To illustrate, for $l=1$, the best 20\% of the matchups are achieved with the distance between partitions does not exceeding (roughly) 9\% for all four parameters simultaneously.}
    \label{C4PM-per-quintil}
\end{table}

%Therefore, the $l$-th experiment of this test consists of determining the $RMSE$ of the quintiles of matchups emanating from the 2PM and C4PM runs when they are initialized taking as input data \( R^{*}_l\) and \( r^{*}_l\) respectively. 

% Analogous analysis can be done on subsequent quintiles to reach similar conclusions, but keeping in mind that the $RMSEs$ get closer as the quintiles advance. The additional and more visual presentation of the evolution of the $RMSEs$ values against the percentage amount of matchups for each experiment ~--~ seen in Figure \ref{fig:$RMSE$_N2} ~--~ confirms this. The image demonstrates, in particular, that C4PM did better in forming matches compared to 2PM, regardless of the percentage of the total amount of matches considered or, in other words, C4PM matched more correctly. 

\begin{figure}[ht!]
    
    \centering
    \begin{subfloat}{
        \centering
        \includegraphics[width=6cm]{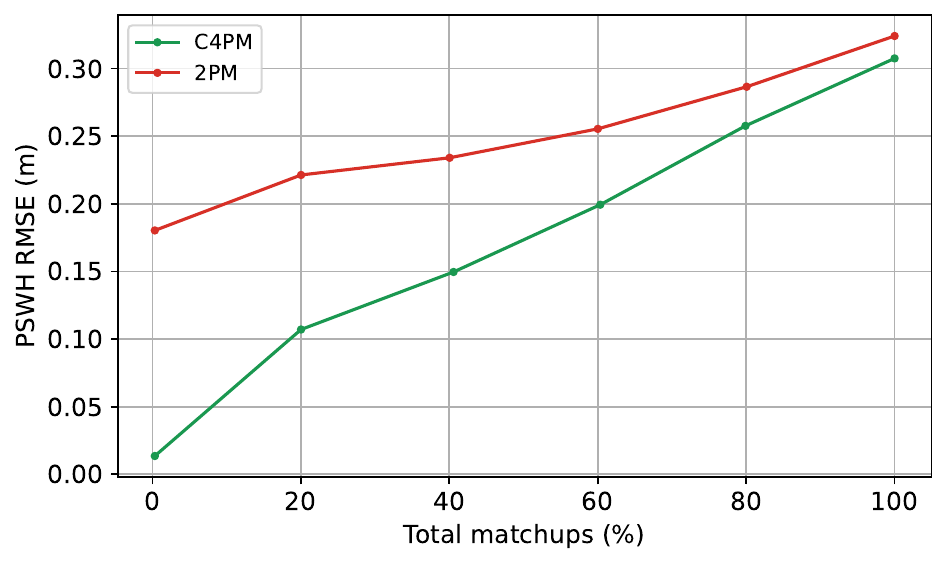}
        \label{subfig:ca_$RMSE$_hs}}   
        \caption{}
    \end{subfloat}
    %\hfill    
    \hspace*{0.5cm}
    \begin{subfloat}{
        \centering
        \includegraphics[width=6cm]{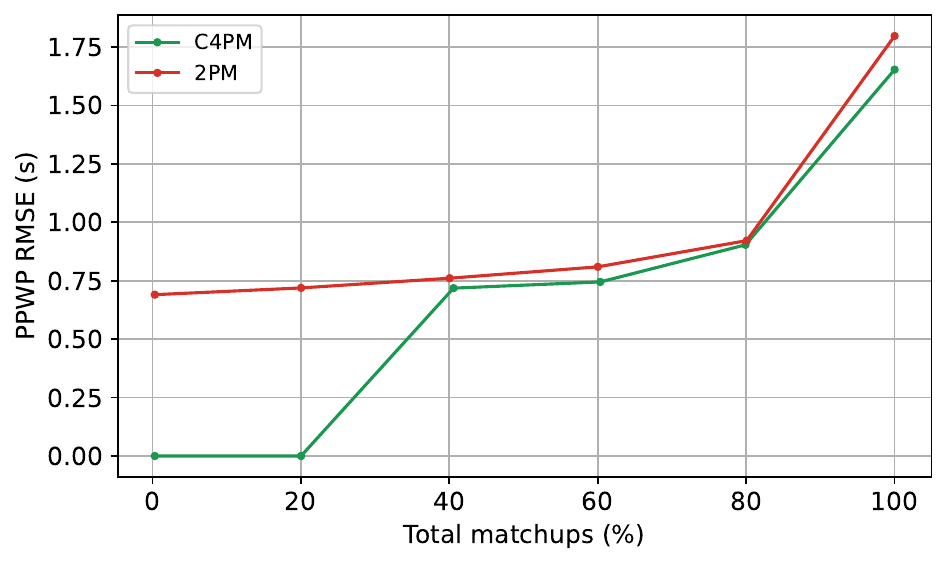}
        \label{subfig:ca_$RMSE$_tp}}   
        \caption{}
    \end{subfloat}
    
    \centering
    \begin{subfloat}{
        \centering
        \includegraphics[width=6cm]{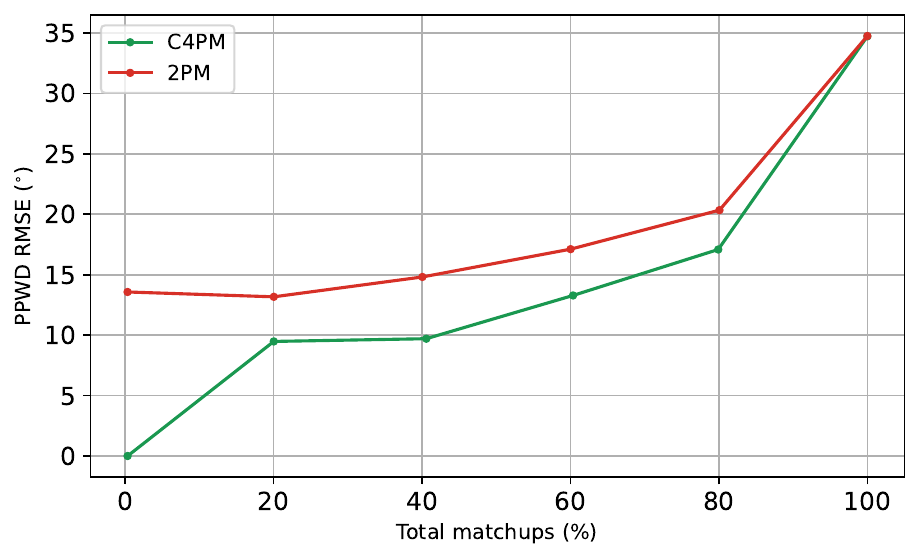}
        \label{subfig:ca_$RMSE$_dp}}    
        \caption{}
    \end{subfloat}
    %\hfill    
    \hspace*{0.5cm}
    \begin{subfloat}{
        \centering
        \includegraphics[width=6cm]{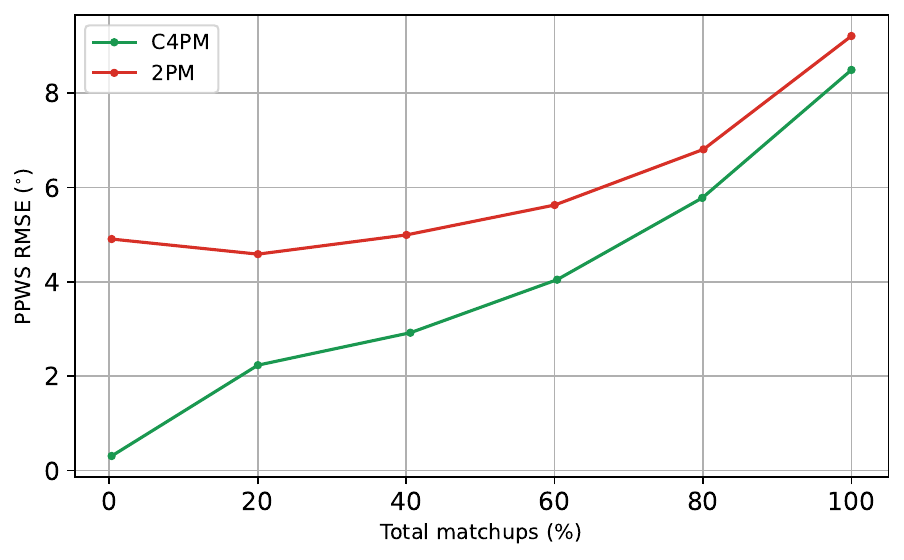}
        \label{subfig:ca_$RMSE$_sp}}        
        \caption{}
    \end{subfloat}
        
    \caption{$RMSEs$: (a) $PSWH$, (b) $PPWP$, (c) $PPWD$ and (d) $PPWS$.}
    \label{fig:$RMSE$_N2}
\end{figure}

An analogous analysis can be performed for subsequent quintiles, yielding similar conclusions, while noting that the $RMSE$ values become increasingly similar as the quintiles progress. 
The additional, more visual representation of the evolution of $RMSE$ values against the percentage of matchups for each experiment~---~shown in Figure~\ref{fig:$RMSE$_N2}~---~further corroborates this observation. 
In particular, the figure demonstrates that C4PM outperformed 2PM in forming matches, regardless of the percentage of total matches considered. 
In other words, C4PM consistently produced more accurate matchups.

\begin{figure}[ht!]
\centering
\subfigure[]{%
\label{subfig:xa}\resizebox*{6.5cm}{!}{\includegraphics{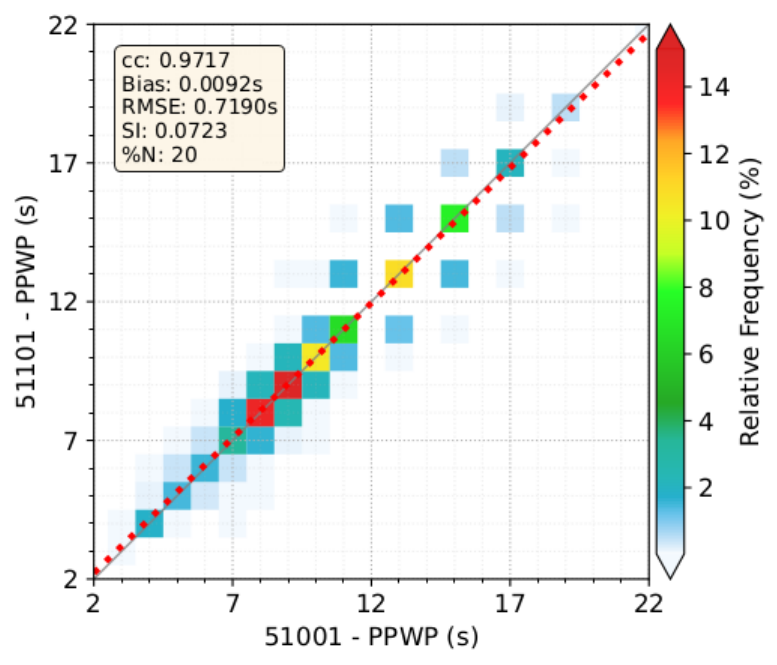}}}
\subfigure[]{%
\label{subfig:xb}\resizebox*{6.5cm}{!}{\includegraphics{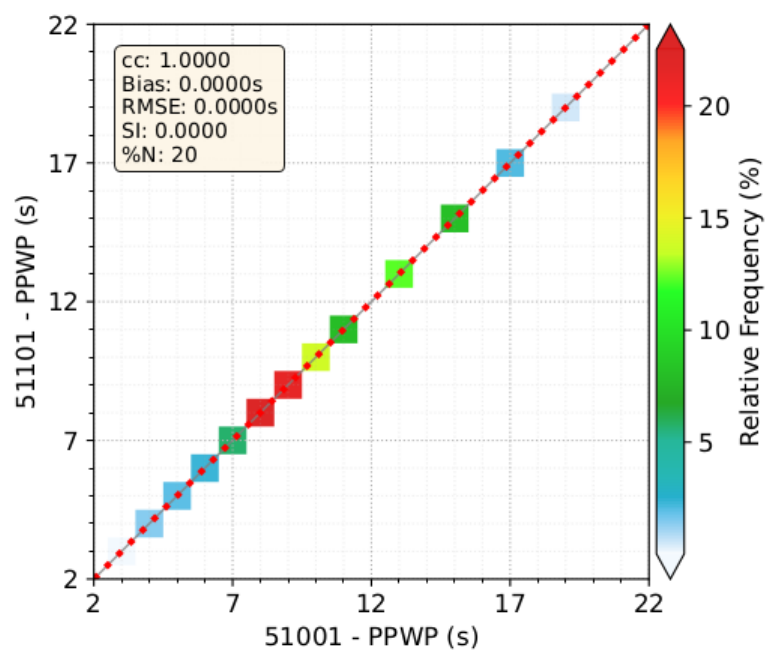}}}

\subfigure[]{%
\label{subfig:xc}\resizebox*{6.5cm}{!} {\includegraphics{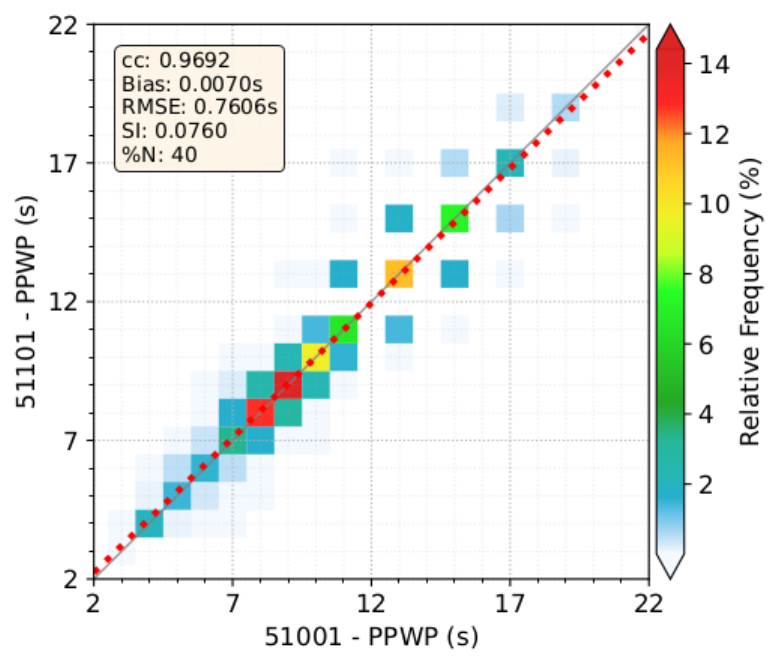}}}
\subfigure[]{%
\label{subfig:xd}\resizebox*{6.5cm}{!} {\includegraphics{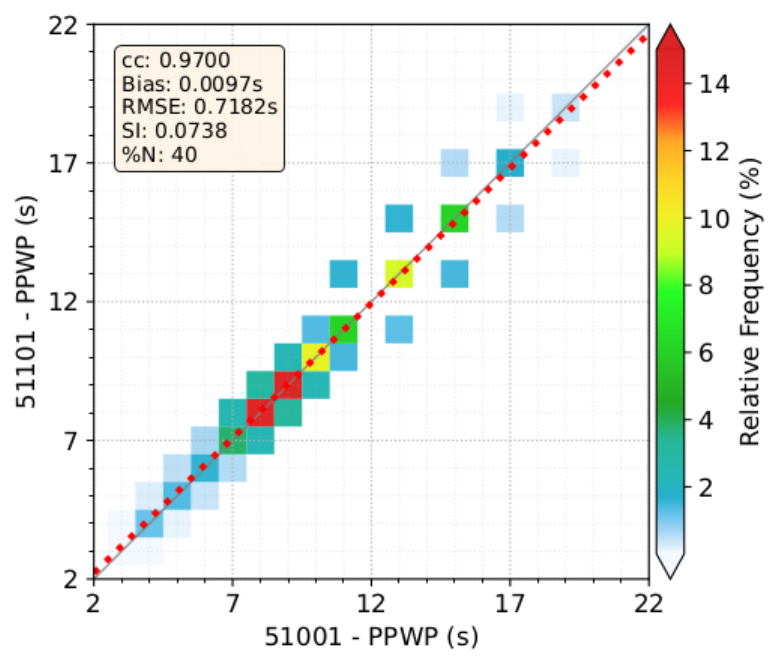}}}

%\subfigure[]{%
%\label{subfig:xe}\resizebox*{4.2cm}{!} {\includegraphics[width=\textwidth]{scatter_tp_q3_hh.pdf}}}
%\subfigure[]{%
%\label{subfig:xf}\resizebox*{4.2cm}{!} {\includegraphics[width=\textwidth]{scatter_tp_q3_ad.pdf}}}

\caption{{PPWP scatter plot for the first two quintiles. Left column, 2PM [ac]; right column C4PM [bd]}. First line is the first quintile [ab]; second line is the second quintile [cd].}
\label{fig:quatile_tp}
\end{figure}

Figure~\ref{fig:quatile_tp} illustrates the PPWP evolution for C4PM and 2PM across the first and second quintiles. 
%The scatter plots provide clear evidence of distinct matching behaviors between the two methods.
In the first quintile, C4PM exhibits a concentrated distribution along the diagonal, indicative of accurate matches and the absence of outliers. 
Conversely, 2PM demonstrates a more dispersed distribution, with significant outliers reflecting lower match quality.
In the second quintile, this trend persists. 
C4PM continues to maintain a narrow distribution along the diagonal, highlighting its ability to generate reliable matches. 
In contrast, 2PM displays a broader distribution accompanied by many outliers, further emphasizing its inferior match quality.
These observations align with the $RMSE$ values reported in Tables~\ref{2PM-per-quintil} and \ref{C4PM-per-quintil}. 
A lower global $RMSE$ generally corresponds to improved match quality and fewer outliers, particularly under stricter matching conditions. 
The consistent superior performance of C4PM in both quintiles underscores its robustness and reliability in producing accurate matches, even under challenging scenarios.

% adicionando figuras para avaliacao

\subsection{Two settings confronted}

% Now, the results obtained from running the 2PM and C4PM algorithms on somewhat broad input data are compared. The test is not a search for equivalent configurations between the two methods, if that is even possible; rather, it is an attempt to understand the behavior of C4PM and the results it provides. Among other things, this also highlights one of the capabilities of the C4PM, which allows, depending on the application, type of measurement or desired accuracy, to control - independently - discrepancies between cross-assigned partitioned integrated wave parameters of different natures. 

The results obtained from running the 2PM and C4PM algorithms on relatively broad input data are compared. 
This test does not aim to identify equivalent configurations between the two methods~---~if such equivalence is even possible~---~but rather to analyze the behavior of C4PM and the results it produces. Notably, this comparison underscores one of C4PM's key capabilities: its ability to independently control discrepancies between cross-assigned partitioned integrated wave parameters of different natures, depending on the application, type of measurement, or desired level of accuracy.

\begin{figure}[ht!]
\centering
\subfigure[]{%
\label{subfig:9a}\resizebox*{4.2cm}{!}{\includegraphics{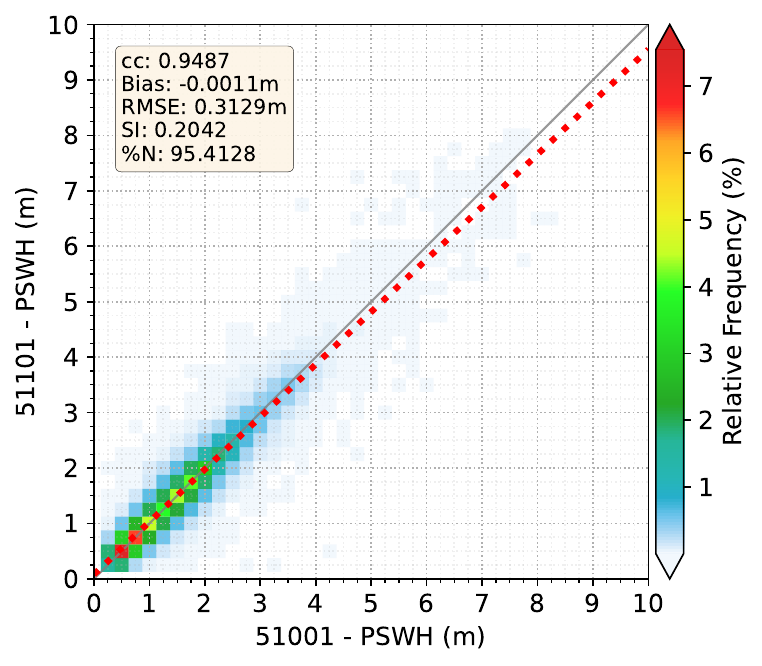}}}
\subfigure[]{%
\label{subfig:9b}\resizebox*{4.2cm}{!}{\includegraphics{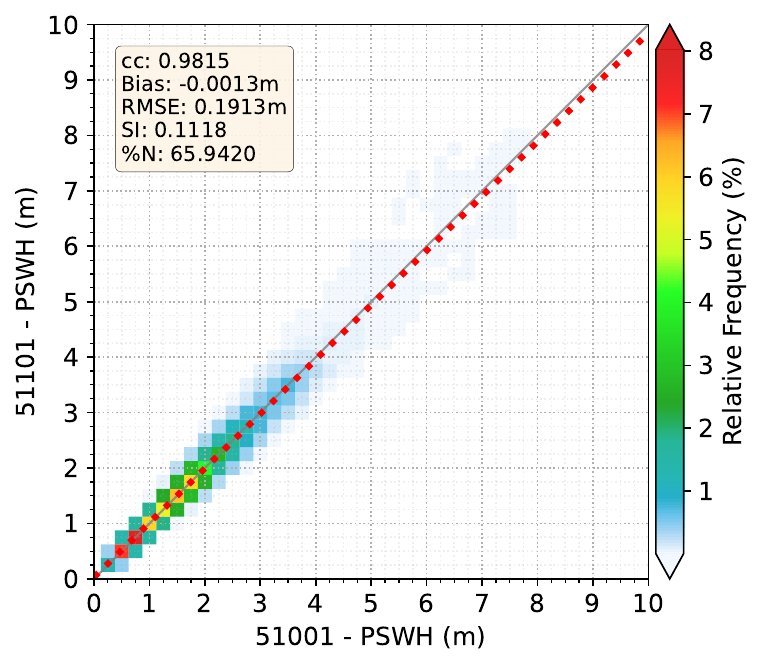}}}

\subfigure[]{%
\label{subfig:9c}\resizebox*{4.2cm}{!} {\includegraphics{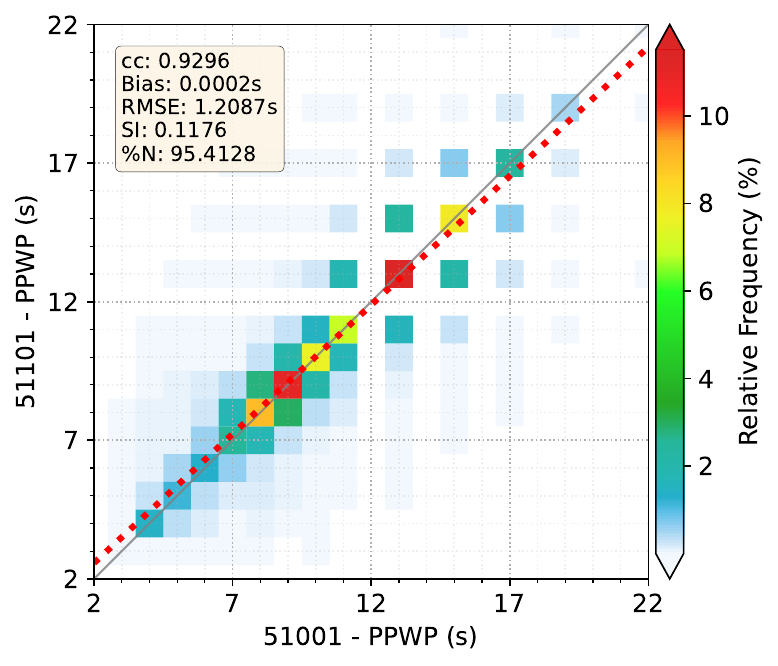}}}
\subfigure[]{%
\label{subfig:9d}\resizebox*{4.2cm}{!} {\includegraphics{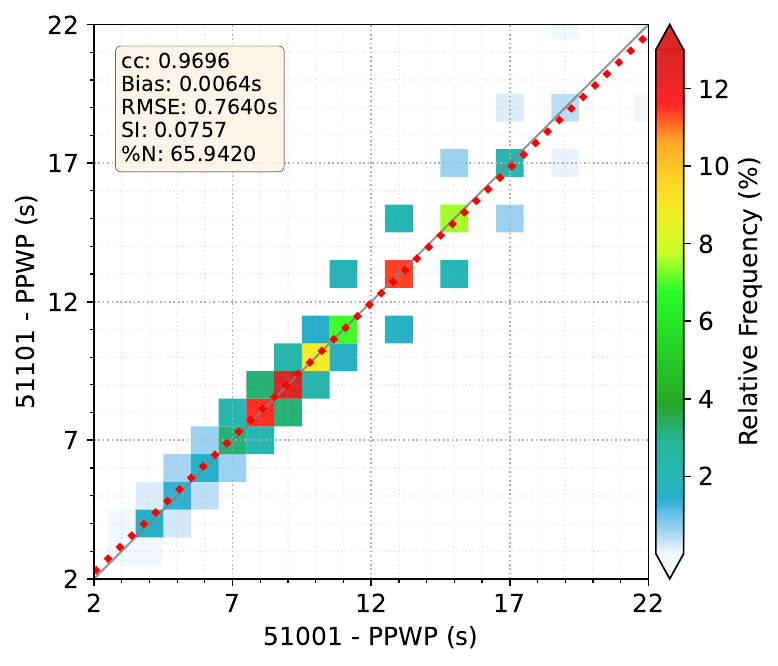}}}

\subfigure[]{%
\label{subfig:9e}\resizebox*{4.2cm}{!} {\includegraphics[width=\textwidth]{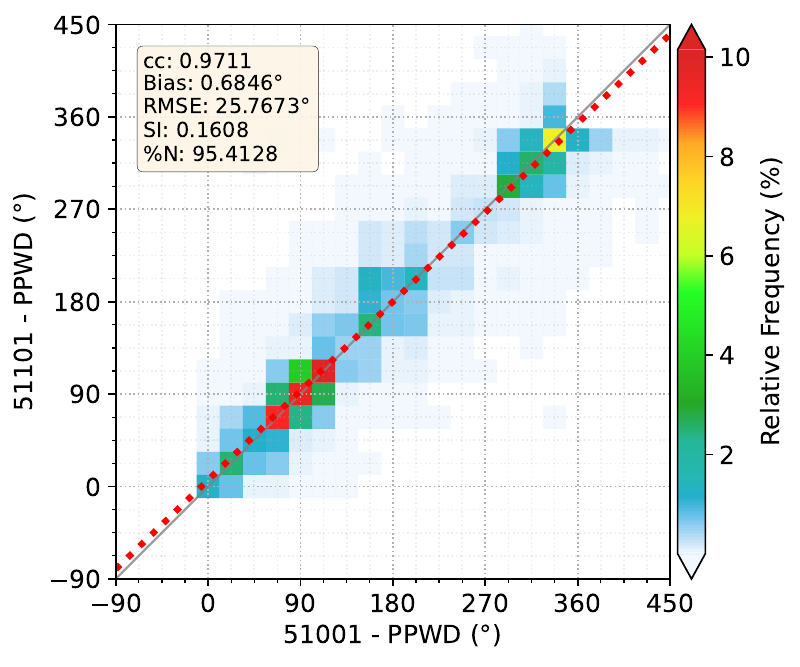}}}
\subfigure[]{%
\label{subfig:9f}\resizebox*{4.2cm}{!} {\includegraphics[width=\textwidth]{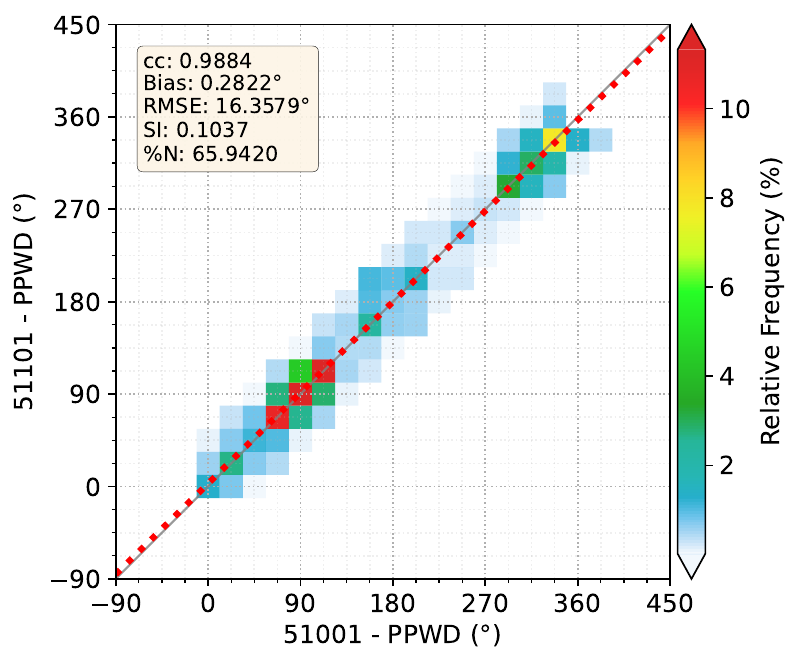}}}

\subfigure[]{%
\label{subfig:9g}\resizebox*{4.2cm}{!} {\includegraphics[width=\textwidth]{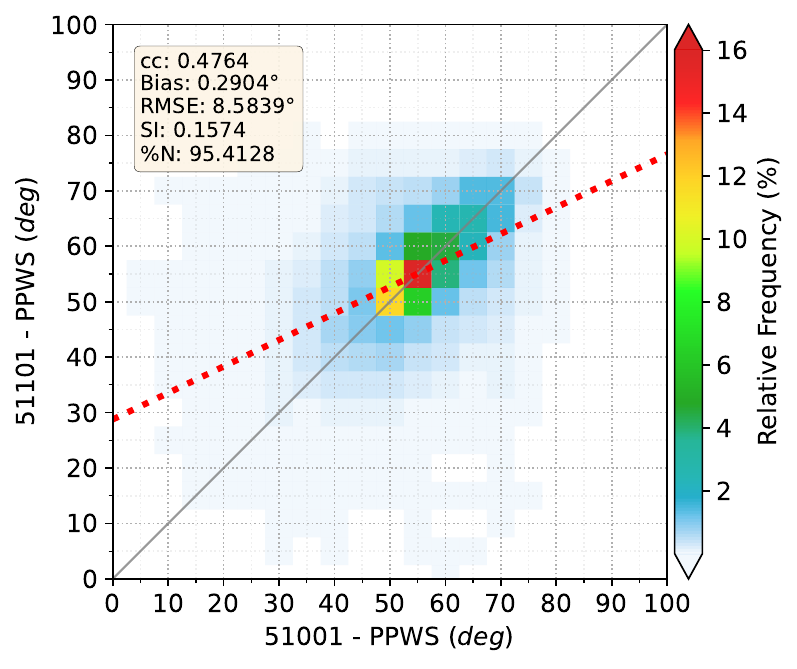}}}
\subfigure[]{%
\label{subfig:9h}\resizebox*{4.2cm}{!} {\includegraphics[width=\textwidth]{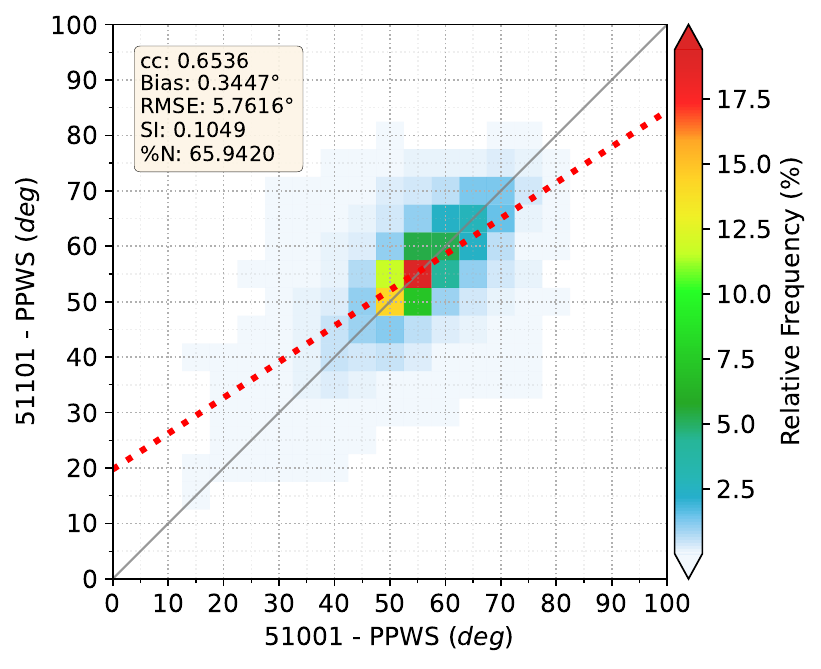}}}
\caption{{Scatter plots between buoys 51001 and 51101 for $PSWH$, $PPWP$, $PPWD$ and $PPWS$, respectively from top to bottom. Left column: 2PM; right column: C4PM.}}
\label{fig:fig6}
\end{figure}

Following \cite{hasselmann1996}, \(R = 0.75\) is used as the critical input value for the 2PM algorithm, while the control vector \(\mathbf{c} = (0.2,0.3,0.2,0.6)\) is applied as input to the C4PM algorithm. 
This control vector imposes the following limits on parameter discrepancies within a matchup: the smallest significant wave height must be at least $(1-0.2) = 0.8$  of the largest significant wave height in the same matchup; similar limitations apply to discrepancies in peak wave periods and peak directional spreads. 
Additionally, cross-assigned peak wave directions cannot differ by more than $0.2 \times 180^{\circ} = 36^{\circ}$.
Figure~\ref{fig:fig6} illustrates the results of this comparison. 
Both methods exhibit negligible bias, high correlation coefficients, and low $RMSE$ for all partitioned integrated wave parameters. 
However, closer examination of Figure~\ref{fig:fig6}[aceg], which displays the 2PM results, reveals the presence of outliers, indicating the occurrence of unlikely matchups. 
Such outliers, while rare, may pose challenges in applications such as data assimilation. 
Notably, these outliers persist even when the critical value for 2PM is significantly reduced (plots not shown).
In contrast, Figure~\ref{fig:fig6}[bdfh], presenting the C4PM results, shows matchups tightly clustered around the 1:1 line, indicating the absence of outliers. 
This result highlights the effectiveness of C4PM in controlled matchup formation, further demonstrating its superiority in avoiding spurious matches.

\section{Summary and Conclusions} \label{SC}

Cross-assignment of wave spectra represents a combinatorial optimization challenge in physical oceanography. 
The core question is how to determine the best matching between partitions of wave spectra from different sources, specifically those whose oceanic characteristics exhibit the highest level of concordance. Addressing this problem requires a reliable method for measuring the similarity between partitions and a robust criterion to determine when pairs of partitions should form the cross-assignment.

% antigo ********
%In this work, we propose a novel spectral distance that incorporates four integrated wave parameters and introduce a non-standard approach to matchup formation, termed the Controlled Four-Parameter Cross-Assignment Problem. 
%To solve this problem, we developed and implemented the Controlled Four-Parameter Method (C4PM), a computational routine designed to perform the cross-assignment process. 
%C4PM offers a high degree of customizability, enabling the assignment of distinct weights to the integrated wave parameters during the distance calculation, while also allowing for \textit{a priori} control of discrepancies among the corresponding parameters throughout the cross-assignment process.
% ********

% minha sugestao ********
This work proposes a novel cross-assignment methodology based on a spectral distance that incorporates four integrated wave parameters, termed the Controlled Four-Parameter Method (C4PM), which offers a high degree of customizability, enabling the assignment of distinct weights to the integrated wave parameters during the distance calculation while also allowing for \textit{a priori} control of discrepancies among the corresponding parameters throughout the cross-assignment process.
% ********

To evaluate the proposed method, we compared C4PM with an existing approach based on a two-parameter spectral distance (denoted as 2PM). 
Thousands of data points from two buoys located 13~km apart were analyzed. 
The experiments explored the progression of cutoff values for each technique and their qualitative impact on the resulting groups of matchups. 
The results consistently demonstrated the superiority of C4PM over 2PM in nearly all aspects.

The weighted semimetric underlying C4PM proved to be well-scaled, providing superior capability in identifying matches with highly concordant oceanic features. 
In contrast, the spectral distance employed in 2PM exhibited limitations, allowing for the formation of mismatched pairs in numerous cases. 
When operating with very strict cutoff values, C4PM identified only perfect matches, i.e., measurements from nearly identical pairs of partitions, while 2PM produced a significant proportion of relatively poor matches even within the same cutoff range.

Further analysis, which divided the total set of possible matches into quintiles based on cutoff values, revealed that C4PM consistently outperformed 2PM across all groups, particularly in the first three quintiles. 
The $RMSE$ values achieved by C4PM for each partitioned integrated wave parameter were consistently lower than those of 2PM, highlighting the accuracy of its matchups. 
This distinction was further underscored by the presence of outliers in the 2PM results, which persisted even when stricter cutoff values were applied. 
These outliers, indicative of poorer match quality, were absent in the C4PM results, demonstrating the robustness of its controlled matchup formation.

C4PM's ability to assign specific weights and variability limits to each parameter provides a level of flexibility that enables tailored cross-assignments to suit the requirements of different datasets and applications. 
Furthermore, its negligible computational overhead and planned release as an open-source Python package enhance its accessibility and practicality for broader use.
In summary, the results presented here demonstrate significant improvements over traditional spectral distance methods for the cross-assignment of directional wave spectra. 
C4PM is a highly effective tool for wave data assimilation, where precise cross-assignments are critical for improving the reliability and accuracy of forecast models.

\newpage
\appendix

\section{Appendix - Statistics and wave parameters}

Tables \ref{tab:stats} and \ref{tab:table2} provide the formulas for the statistical parameters and wave parameters, respectively.

\begin{table}[ht!]
  \begin{center}
    
    %\begin{tabular}{@{}cc@{}} % <-- Changed to S here.
    \begin{tabular}{|c c|} % <-- Changed to S here.
    \hline     
      Parameter & Expression\\
    \hline
     
      mean & \begin{math} \overline{X} = \displaystyle\frac{1}{n} \sum_{i=1}^{n} x_{i}\end{math}, \hspace{0.5cm} \begin{math}\overline{Y} = \displaystyle\frac{1}{n} \sum_{i=1}^{n} y_{i} \end{math}\\
      
      \xrowht{40pt}
      bias & \begin{math}\displaystyle\frac{1}{n} \sum_{i=1}^{n} (y_{i}-x_{i})\end{math}\\
      
      \xrowht{40pt}
      root mean square error ($RMSE$) & \begin{math}\displaystyle\sqrt{\frac{1}{n} \sum_{i=1}^{n} (y_{i}-x_{i})^{2}}\end{math} \\
      
      \xrowht{40pt}
      Scatter index (SI) & \begin{math}\displaystyle\frac{\mathrm{(RMSE)}}{\overline{X}}\end{math}\\
      
      \xrowht{40pt}
      Pearson correlation coefficient ($cc$) & \begin{math}\displaystyle\frac{\sum_{i=1}^{n} (x_{i}-\overline{X})(y_{i}-\overline{Y})}{\sqrt{\sum_{i=1}^{n} (x_{i}-\overline{X})^2} \sqrt{\sum_{i=1}^{n}(y_{i}-\overline{Y})^2}}\end{math}\\

      \hline      
      \end{tabular}
      \caption[Statistical parameters.]{Statistical parameters.}
      \label{tab:stats}
  \end{center}
\end{table}

\begin{table}[ht!]
  \begin{center}    
    \begin{tabular}{|c c|} % <-- Changed to S here.
     \hline
      Parameter & Expression\\
     \hline
      $SWH$ & \begin{math}\displaystyle 4 \sqrt{\int_0^{2\pi} \int_0^\infty S(f, \theta) \, df \, d\theta } \end{math}\\
      
      \xrowht{40pt}
      $PWP$ & \begin{math}\displaystyle \frac{1}{\text{argmax}(S(f))} \end{math}\\
      
      \xrowht{40pt}
      $PWD$ & \begin{math}\displaystyle \text{atan2}\left(\frac{\int_{0}^{2\pi} S(f_p, \theta) \sin(\theta) \, d\theta}{\int_{0}^{2\pi} S(f_p, \theta) \cos(\theta) \, d\theta} \right) \end{math} \\
      
      \xrowht{55pt}
      $PWS$ & \begin{math}\displaystyle \sqrt{2 \left( 1 - \sqrt{ \left( \frac{\int_{0}^{2\pi} S(f_p, \theta) \cos(\theta) \, d\theta}{\int_{0}^{2\pi} S(f_p, \theta) \, d\theta} \right)^2 + \left( \frac{\int_{0}^{2\pi} S(f_p, \theta) \sin(\theta) \, d\theta}{\int_{0}^{2\pi} S(f_p, \theta) \, d\theta} \right)^2 } \right) } \end{math}\\
            
              \hline      
      \end{tabular}
      \caption[Wave parameters.]{Wave parameters. $S(f,\theta)$ represents the directional wave spectrum.}
      \label{tab:table2}
  \end{center}
\end{table}

\newpage

% ***************************************************************

\bibliographystyle{agsm} 
\bibliography{references}

\end{document}